\documentclass[trackchanges,twocolumn]{aastex7}
\usepackage{amsmath}
\defcitealias{passmark5995wx}{PassMark Software 2025}
\defcitealias{Hitomi2018}{Hitomi Collaboration et al. 2018}

\begin{document}

\title{$\tt{XFit}$: Global Optimization and Degeneracy Mapping in X-ray Spectral Modeling}

\author[orcid=0009-0002-3696-7339,gname=Austin,sname='MacMaster']{MacMaster A.}
\affiliation{Department of Physics and Astronomy, University of Manitoba, Winnipeg, MB R3T 2N2, Canada}
\email[show]{austin.macmaster@gmail.com}  

\author[orcid=0000-0003-2953-2054,gname=Adam, sname='Rogers']{Rogers A.} 
\affiliation{Department of Physics and Astronomy, Brandon University, Brandon, MB R7A 6A9, Canada}
\email{rogersa@brandonu.ca}

\author[gname=Jason, sname='Fiege']{Fiege J.} 
\affiliation{Department of Physics and Astronomy, University of Manitoba, Winnipeg, MB R3T 2N2, Canada}
\email{fiege@physics.umanitoba.ca}

\author[gname=Rebecca, sname='Man']{Man R.} 
\affiliation{Department of Physics and Astronomy, University of Manitoba, Winnipeg, MB R3T 2N2, Canada}
\email{manb@myumanitoba.ca}

\author[orcid=0000-0001-6189-7665,gname=Samar, sname='Safi-Harb']{Safi-Harb S.} 
\affiliation{Department of Physics and Astronomy, University of Manitoba, Winnipeg, MB R3T 2N2, Canada}
\email{samar.safi-harb@umanitoba.ca}

\begin{abstract}

\noindent The standard approach to modeling X-ray spectral data relies on local optimization methods, such as the Levenberg-Marquardt algorithm. While effective for simple models and speedy spectral fitting, these local optimizers are prone to becoming trapped in local minima, particularly in high-dimensional or degenerate parameter spaces, and typically require extensive user intervention. In this work, we introduce $\tt{XFit}$, a global optimization method for fitting X-ray data, which makes extensive use of the \textit{Ferret} evolutionary algorithm. $\tt{XFit}$ enables automated exploration of complex parameter spaces, efficient mapping of confidence intervals, and identification of degenerate solutions that may be overlooked by local methods. We demonstrate the performance of $\tt{XFit}$ using two representative X-ray sources: the Central Compact Object in Cassiopeia~A and the supernova remnant G41.1--0.3. These examples span both low- and high-dimensional models, allowing us to illustrate the advantages of global optimization. In both cases, $\tt{XFit}$ produces solutions that are consistent with or improve upon those found with traditional methods, while also revealing alternative fits or degenerate solutions within statistically acceptable confidence levels. The automated mapping of parameter space offered by $\tt{XFit}$ makes it a powerful complement to existing spectral fitting tools, particularly as models and data quality become increasingly complex. Future work will expand the application of $\tt{XFit}$ to broader datasets and more physically motivated models.

\end{abstract}


\section{Introduction\label{sec:intro}}
Astrophysical phenomena are ideal laboratories for studying the myriad high-energy emission processes capable of producing X-ray radiation. Such processes include the synchrotron emission emanating from compact objects, pulsar wind nebulae, and shocked plasma in supernova remnants \citep[SNRs;][]{2008ARA&A..46...89R}; the bremsstrahlung, collisional excitation, and recombination radiation in hot, optically thin plasmas of stellar coronae, clusters of galaxies, and SNRs \citep{Heuer2021}; electron scattering in the accretion disks of neutron stars (NSs) and black holes (BHs); and photoionization in X-ray binaries and active galactic nuclei \citep{2008AIPC..983..171K,Mukai2003}. Additionally, X-rays can penetrate deeply into gas and dust surrounding regions of star formation \citep{Raga2002}. In cosmology, X-rays are used to study the photons emitted by the cosmic microwave background as they are scattered by relativistic electrons \citep{Celotti2001}, and large-scale distance estimates can be calculated using elemental abundances inferred from the absorption of X-rays in the interstellar medium \citep[ISM][]{Silk1978}. Plasma interactions in clusters of galaxies can reach temperatures upwards of 10-100 MK (mega-Kelvin), producing thermal X-rays used to estimate baryonic matter densities in cosmological models \citep{Weinberg2013}. Recent studies of the multi-messenger event GW170817 \citep{Abbott2017,Abbott2017-2} are also demonstrating the relevance of X-rays as an electromagnetic counterpart to gravitational wave astronomy in investigating the evolution of merging NS-NS, NS-BH, and BH-BH binary systems in the short gamma-ray burst and kilonova remnant phases \citep{2018ApJ...862L..19N,Safi-Harb2019,Troja2020,Ren2022,Hajela2022}.

The 1962 launch of a Geiger counter demonstrated the feasibility of detecting X-rays outside Earth's atmosphere, and christened a new era of X-ray astronomy \citep{Giacconi1962}. Subsequently, improvements in spectroscopic observing techniques in the field of X-ray astronomy --- such as CCD detector technology \citep{Brickhouse2000}\footnote{\label{chandrafoot}\url{https://cxc.harvard.edu/proposer/POG/html/chap8.html}}\textsuperscript{,}\footnote{\url{https://www.cosmos.esa.int/web/xmm-newton}\label{xmmfoot}} and microcalorimeter arrays \citepalias{Hitomi2018}\footnote[3]{\url{https://www.isas.jaxa.jp/en/missions/spacecraft/past/hitomi.html}\label{hitomifoot}} --- have led to major improvements in detector sensitivity, reductions in background count rates, as well as the spatial and spectral resolution of energy distributions of photons. As data sets increase in size, detail, and complexity, so too do the models predicting the physical properties inferred by the observed spectra, introducing an ever-increasing demand on the reliability and consistency of the optimization schemes used to fit models to data. 

A common approach to model fitting begins with the selection of an objective or `fitness' function that statistically determines how well a given model $\hat{M}(x_k;\textbf{p})$ predicts the probability of detecting an observed event $M(x_k;\textbf{p})$ across an observation's $k$ spectral energy bins for a given set of independent variables $x_k$ and their measurement error $\sigma({M(x_k))}=\sigma_{M_k}$. Nonlinearities in the parameters \textbf{p} of a chosen model pose challenges for local linear least-squares fitting methods, which become unreliable or inefficient. Furthermore, gradient-based methods require that the objective function be smooth and continuous, and if any of the partial derivatives of the objective function are nonlinear with respect to the model parameters, the space of solutions must be searched for a minimum by iteratively adjusting parameter values based on the topology of the surrounding parameter space until a minimum fit value is obtained. The choice of objective function is problem specific and can depend on multiple factors, such as the total number of counts present in a spectrum. For example, low count rates are typical in X-ray astronomy, where distributions of counts may best be described by Poisson statistics. A commonly used fit statistic in the low-count regime is the Cash statistic \citep{Cash1979} and its variants. For high numbers of counts, the Poisson distribution is well-approximated by a Gaussian distribution, and a minimization of the sum of the weighted squares of the errors between $\hat{M}(x_k;\textbf{p})$ and $M(x_k;\textbf{p})$ is performed over n energy bins as shown in the objective function $\chi^2$ of Equation \ref{eq:chisquare}.

\begin{align}
\chi^2(\textbf{p})   = & \hspace{0.2em} \sum_{k=1}^n \bigg[ \frac{M(x_k) - \hat{M}(x_k;\textbf{p})}{ \sigma_{M_k} } \bigg]^2 \nonumber \\ 
  = & \hspace{0.2em} \textbf{M}^T\textbf{W}\textbf{M} - 2\textbf{M}^T\textbf{W}\hat{\textbf{M}} + \hat{\textbf{M}}^T\textbf{W}\hat{\textbf{M}},
\label{eq:chisquare}
\end{align}

\noindent where \textbf{M} and $\hat{\textbf{M}}$ are the matrix forms of the observed events and model, respectively, and \textbf{W} is the inverse of the error covariance matrix. 

The XSPEC\footnote[4]{\url{https://heasarc.gsfc.nasa.gov/xanadu/xspec/}\label{xspecfoot}} \citep{Arnaud1996} fitting package contains multiple minimization methods for fitting models with non-linear parameters. XSPEC’s default optimizer, the modified Levenberg-Marquardt \citep[LM;][]{Levenberg1944} algorithm based on the CURFIT routine from Bevington \citep{Bevington2003}, is a commonly used approach to model fitting in high-energy astrophysics. The LM algorithm (LMA) is designed to minimize the objective function by updating the model parameters by a perturbation \textbf{h} that is scaled by a damping coefficient $\lambda$, as shown in Equation \ref{eq:LMA}. 

\begin{equation}
\Big[\textbf{J}^T\textbf{W}\textbf{J}+\lambda\textbf{I}\Big]\textbf{h}=\textbf{J}^T\textbf{W}(\textbf{M}-\hat{\textbf{M}})
\label{eq:LMA}
\end{equation}

\noindent  where $\textbf{J}$ is the Jacobian matrix.

The LMA approaches the optimization differently depending on the local geometry of the parameter space surrounding the current search point. The LMA starts with an initial guess for the parameters $\textbf{p}_0$ and $\lambda_{i=0}>>1$. Far from the putative minimum, $\lambda$ is large and the LM algorithm behaves like a gradient-based optimizer, simultaneously adjusting the step size of the first derivatives of each parameter and orienting the search in the direction of steepest descent. As the search approaches a minimum, gradients tend toward zero as the topology of the parameter space becomes flattened. Small step sizes in this region come at the cost of computational efficiency, while large step sizes result in a loss of precision and increased risk of moving further from the solution by stepping over the minimum. The LMA circumvents this issue by iteratively updating $\lambda$ to find perturbations in the model parameters that contribute to an overall decrease in the objective function at successive steps. If a step leads to improvement in the fit, $\chi^2(\textbf{p}+\textbf{h}) < \chi^2(\textbf{p})$ and the objective function is then evaluated at a trial next-step using a reduced $\lambda_{i+1}<\lambda_i$. If $\chi^2(\lambda_{i+1}) \approx \chi^2(\lambda_i)$, the quadratic approximation is considered valid and $\lambda$ is decreased. Otherwise $\lambda$ is increased. As $\lambda \rightarrow0$, Equation \ref{eq:LMA} reduces to the Gauss-Newton method of minimization, where $\chi^2$ is Taylor expanded to second-order and approximated as a parabolic surface.

The LMA is an example of a local, single-search-agent, gradient-based optimizer. Such optimizers evaluate very quickly, perform well on simple, low-dimensional problems, and return a fitness value arbitrarily close to the minimum in the convex parameter region. However, one can imagine many hypersurfaces for which gradient-based optimization methods would get stuck in local minima, missing other potentially better or degenerate solutions wherein different sets of parameters evaluate to fitness values within 3$\sigma$ of the global-best solution.

Models that are fully or partially based on data tables may not be smooth or continuous in the model parameters or their derivatives, and noise may look like many small local minima for an optimizer, resulting in poorly defined gradients. Additionally, a steepest-descent gradient-based algorithm is incapable of exploring other regions a of a parameter space once a local minimum is found, since a step size in any direction results in a poorer objective function value. For these reasons, it is common practice for researchers to use information collected through simulations or from the literature for similar astrophysical scenarios to make an educated `best initial guess' of a reasonable starting point for the search. For complicated problems, the final result may be sensitive to this initial condition in practice, since local optimizers are prone to become trapped in local minima. Other practices, such as reducing the hypervolume of the search space by holding particular model parameters constant and performing a local search within a subspace, are also common. However, not all problems can be solved by sequentially optimizing arbitrarily chosen subspaces, which are still prone to becoming trapped in local minima or potentially introducing bias into the results of the fit. 

A single-agent gradient-based optimization algorithm’s reliance on differentiable objective functions and arbitrary user-input imposes limits on researchers’ ability to fit models to data containing both systematic uncertainty and noise. This motivates the question as to whether novel approaches such as global optimization methods represent suitable alternatives that can improve the consistency with which physically interesting solutions are found, while also reducing potential biases introduced by the aforementioned conventional approaches to model fitting. Global Bayesian samplers interfaced with XSPEC provide a powerful alternative to local optimization for fitting models to X-ray spectra. The Bayesian X-ray Analysis (BXA) software \citep{2014A&A...564A.125B,2021JOSS....6.3001B} is a Bayesian-inference front-end that connects XSPEC models to the ``UltraNest'' nested sampling engine  \citep{2021JOSS....6.3001B} \citep[originally MultiNest, ][]{2009MNRAS.398.1601F} that demonstrates improvements over a number of limitations typically associated with the use of standard Markov chain Monte Carlo (MCMC) methods in complicated likelihood landscapes. MultiNest uses multi-ellipsoid clustering to track and apply weights to separate local maxima in a fully Bayesian manner, while UltraNest adds region-based metrics in the evidence and posterior, enabling Bayesian inference in multi-modal problems.

The computational cost required to compute evidence and posterior estimates using nested sampling scales linearly with the number of live points and at least quadratically with dimension \citep{2023StSur..17..169B}. In applications where the target is to map the profile likelihood rather than compute Bayesian evidences, differential evolution–based scanners have been found to outperform MultiNest and several variants of MCMC in terms of computational cost for models with $\gtrsim 10$ free parameters \citep{2017EPJC...77..761M}. Furthermore, optimizers that utilize evolutionary strategies are particularly effective at identifying and mapping multi-modal structures of high-likelihood regions in parameter space for high-dimensional problems where evidence computation would be prohibitively expensive for large data sets. Section \ref{sec:xfit} will provide a broad overview of the global evolutionary optimization methods used to fit spectroscopic models to the two sources in Section \ref{sec:fitting}, and Section \ref{sec:results} describes the analytical methodology used in comparing the results of the optimizations.

\section{XFit: Global Optimization} \label{sec:xfit}

High-dimensional objective functions containing noise, uncertainty, and multiple local optima pose significant challenges to optimization. Problems of this type also represent rich testbeds for determining which characteristics in a given model present difficulties to an optimization algorithm, the exploration of alternative optimization schemes that may perform better on some or all of these characteristics, and identifying robust indicators for the efficiency and efficacy of said optimizers. Genetic Algorithms were first described by \cite{Holland1975} and operate in analogy to the principle of biological evolution \citep{Goldberg1989}. Examples include the PIKAIA algorithm \citep{Charbonneau1995}, which was offered as an option in previous versions of $\tt{XSPEC}$ \citep[v.11.3;][]{Arnaud1996}. PIKAIA implements two basic genetic operations acting on a population of strings encoded using a decimal alphabet: a uniform one-point crossover operator and uniform one-point mutation operator. However, this simple genetic algorithm was found to be inefficient at optimizing spectral model parameters. As mentioned in the $\tt{XSPEC}$ v11.3 manual, “tests with 3 free parameters show that thousands of generations are required to converge on the correct fit”. This is evidence that the fitting problem is difficult for a simple GA with limited operators and precision. The “genetic” option in $\tt{XSPEC}$ was later removed in successive versions as it was a seldom-used fitting method in that package. In this paper, we introduce $\tt{XFit}$, an X-ray spectral fitting package written in MATLAB \citep*{MATLAB} which uses the Qubist Global Optimization Toolbox \citep{fiege-2010} to fit spectral models to X-ray data. The Qubist package includes the Ferret Evolutionary Optimizer which is specially designed to explore the structure of the parameter space by making use of advanced mapping capabilities and the ability to output degenerate solutions where different sets of parameters map to objective function values within 3$\sigma$ (99.7\% confidence) of the global best solution.

While traditional GAs use a binary representation of the objective function parameters \citep{Holland1975}, evolutionary optimizers (EOs), such as Ferret, use real-valued parameters that are acted upon by the algorithm’s various operators. EOs use a stochastic search approach to explore the parameter space in a random, yet directed manner, and like GAs, borrow from the principles of natural selection to search for a global minimum while also maintaining diversity in a population of parameter sets each mapping to a fitness value. This diversity gives the EO the ability to search for degenerate solutions while simultaneously mapping regions of the parameter space satisfying a user-specified statistical fitness criterion. Each set of parameter values is encoded in an array that represents a ‘member’ of a population that is transformed at each time-step or ‘generation’ of the algorithm by operators such as the crossover, mutation, and selection operators.

A tournament-based selection operator assigns copies to each member of the population to pass on to the next generation. The number of copies assigned to each member is calculated based on the value of the individual’s objective function. To ensure the fitness monotonically decreases, a user-defined fraction of the most-fit solutions or `elites' are guaranteed to pass directly to the next generation without modification. Crossover operators select members to swap a single parameter or combinations of parameters as `building blocks', and a secondary geometric crossover operator mixes sets of parameters by treating the two sets of parameters as coordinates in a vector space and taking the stochastic average between them to generate new parameter sets known as `offspring’. The array elements of each individual are acted upon by a mutation operator, which perturbs the element by a value chosen from an empirical distribution that co-evolves with the population to maximize success probability. The mutation operator is also highly directed, making use of the semi-local distribution of the population and their fitness values. The probability of an element being selected for mutation as well as the standard deviation of perturbation in both mutation and geometric crossover operations can be set by the user and even optimized by the algorithm as the search progresses. With the iteration of these three basic processes, the population moves toward individuals of superior fitness, exploring the parameter space, and converging on the global minimum.

Techniques such as `niching' promote diversity in the population by penalizing solutions that are too similar to each other. This diversity drives the thorough exploration of a parameter space around isolated solutions, allowing the algorithm to potentially discover and map distinct islands of solutions, a process that is crucial for multi-objective objective functions. Niching is also useful for mapping confidence intervals as an $\tt{XFit}$ run proceeds and helps avoid premature convergence. Another of Ferret’s useful features is its `linkage learning' algorithm, which attempts to monitor nonlinearities in the problem. If nonlinearities are found, Ferret will attempt to reduce the parameter space into separate subspaces (when possible), which can then be optimized quasi-independently. 

$\tt{XFit}$ makes use of a MATLAB MEX interface to interface with the existing library of spectral models that are distributed with the $\tt{XSPEC}$ code, written in C and FORTRAN. The flexibility of this approach allows us to include new models that are frequently written and updated by the $\tt{XSPEC}$ user base. Moreover, fitting with the Qubist toolbox is an automatic procedure that does not require extensive user interaction common to local optimizers, such as guessing a best initial starting point or freezing parameters and fitting over the resulting lower-dimensional subspace. The global nature of evolutionary optimizers allows the process to proceed with minimal external influence, reducing the chances of introducing bias into the fitting procedure.

Global optimization methods such as Ferret often require a greater number of calls to the objective function per evaluation step than the local LMA, although Ferret's computations are easily parallelized to make use of multi-core computers or clusters. Despite the sometimes large number of evaluations required to converge to a solution, the advantages of global approaches become obvious in mapping degeneracies and thoroughly exploring the parameter space of complex models. $\tt{XFit}$ provides a new method that is capable of finding multiple degenerate solutions with model fit statistics that predict the observed data better than the standard approach while automatically mapping confidence intervals during a run. Although convergence to an optimal solution with the EO may be slower than with the LM method, it is more likely to find solutions that are difficult to find with the usual manual search of parameter subspaces as models increase in size and complexity.

$\tt{XFit}$ is an exploration tool that produces a robust set of optimal solutions even when exploring high-dimensional parameter spaces, and illustrates subtle, nontrivial behaviors of models over large regions of parameter space, saving valuable human time and effort in exchange for computing cycles. We see $\tt{XFit}$ as a valuable tool that augments the traditional approach, overcomes many limitations inherent in local optimizers, and usually finds better solutions. 

As a proof of concept, we apply $\tt{XFit}$ to two representative X-ray datasets. The CCO in Cassiopeia A and the western lobe of the SNR G41.1$-$0.3. These examples were selected to demonstrate the application of $\tt{XFit}$ to a low-dimensional blackbody spectral model in compact objects, and a high-dimensional spectral model describing thermal emission from an ejecta dominated SNR, respectively. The present study focuses on these two sources for illustrative purposes, and future work will extend the application of $\tt{XFit}$ to larger source samples and more physically motivated models.

\section{Spectral Fitting} \label{sec:fitting}
SNRs come in a variety of morphologies made up of some combination of compact object, wind nebula, and shell of shocked and heated material ejected into the interstellar medium. SNRs emit radiation through a myriad of processes with dependencies on elemental composition; electron temperature; ionization and recombination timescales; turbulent velocity flows; and non-thermal particle distributions. Accurate measurements of X-ray emission lines are important for estimating plasma temperatures and the abundances of nucleosynthesis products in SNR ejecta. A spectral line is typically characterized by its amplitude, width, and centroid energy. In spectroscopy, line amplitudes or emissivities are used to determine elemental abundances contained within a source. Lines are attributed a natural width due to quantum mechanical uncertainty, but deviations from this natural line width can also be used to measure the thermal motion of the plasma. Additionally, Doppler shifting of line centroids are often used to determine the distributions of velocities within a plasma. Furthermore, several studies \citep[e.g.,][]{Yamaguchi2014, 2023MNRAS.525.6257B} demonstrate a strong correlation between progenitor type, explosion mechanism, and the strong optically-thin emission lines of ejecta (including the Fe-K line centroid) in thermal X-ray spectra of SNRs. Depending on the energy resolution of the observation, multiple species as well as ions of the same species, may contribute to the intensity of a line, implying the possibility of degeneracy in the parameter space of solutions.

Theoretical models aim to predict the count rates observed by a detector as a function of photon energy. The process of spectral fitting can be described by performing a simultaneous regression between the observed and predicted count rates over a range of energy bins. An X-ray telescope's pulse height amplitude (PHA) encodes the photon energy as the integrated charge per pixel of the ``event”, and a background subtraction region can be used to improve the signal-to-noise of the spectrum. An ancillary response file containing information on the energy- and time-dependence of the detector area’s quantum efficiency is multiplied by the model photon spectrum, producing a spectrum that would be seen by a detector with perfect spectral resolution. This spectrum is then multiplied by a response matrix file, which maps from energy space into pulse height, effectively spreading the counts by the detector’s energy resolution and producing the final spectrum.

The $\tt{XSPEC}$ data analysis software utilizes custom made and pre-loaded models to fit to data. The ‘Tuebingen-Boulder ISM absorption’ (TBabs) model is used to account for X-ray absorption due to hydrogen, where the observed intensity $I_{obs}(x_k)$ of the X-ray spectrum of a source with emitted intensity $I_{source}$ is given by Equation \ref{eq:absorb}.

\begin{equation}
I_{obs}(x_k)=e^{-\sigma_{ISM}(x_k)N_H}I_{source}(x_k),
\label{eq:absorb}
\end{equation}

Where N$_{H}$ is the total hydrogen column density in units of cm$^{-2}$. \cite{Wilms2000} provide updated ISM abundance values and ionization cross-sections to TBabs by assuming a default value for molecular hydrogen of 20\%. Photoelectric cross-sections obtained by \cite{Balucinska-Church1992} are used with the TBabs model in the following analysis of the CCO in Cassiopeia A (see \ref{subsec:casa}). To enable direct comparison with earlier work, we fit G41.1--0.3 (\ref{subsec:casa}) using a photoelectric absorption model based on the Wisconsin cross-sections \citep{Morrison1983}, that predicts a model spectrum given by Equation \ref{eq:wabs}.

\begin{equation}
M(x_k;\textbf{p})=e^{-N_{H}\sigma(x_x)},
\label{eq:wabs}
\end{equation}

\noindent where $\sigma(x_k)$ is the photo-electric cross-section neglecting Thomson scattering and the relative abundances are obtained by \cite{Anders1989}.

\subsection{The CCO in Cassiopeia A} \label{subsec:casa}
Cassiopeia A (Cas A) hosts one of the most widely known CCOs, CXOU J232327.9+584842, which was first seen by the Chandra ``first light" observation in 1999 \citep{Tananbaum1999, 2000ApJ...531L..53P}. Spectral and timing analysis by \cite{heyl_2001} revealed that the point source in Cas A shows no signs of pulsations and different spectral properties than other young pulsars such as the Crab pulsar. CCOs commonly reside near the geometric centre of young ($0.3-7$ kyr) SNRs, producing thermal X-ray emission with no radio or optical counterparts. Their spectra are typically well fit to that of a hot blackbody model of temperature $\sim 2 - 6 \times 10^6$ MK \citep{deluca_2017} or to a power law spectrum with a steep photon index \citep{pavlov_2004,pavlov_2001}. Although young NSs are typically expected to produce pulsations, only 3 CCOs have been detected so far as pulsars \citep{2013ApJ...765...58G}. Possible explanations for the lack of pulsations in most CCOs include weak magnetic fields ($< 10^8$ G), uniform surface temperature, or a viewing angle from Earth that prevents us from seeing any pulsations. When the emitting radius is derived from these models, the result is typically on the order of $\sim$ 1 km, much smaller than the $\sim$ 10 km canonical size of a NS \citep{ozel_2016} implying the existence of hotspots on the NS surface that are usually associated with large magnetic fields. \cite{pavlov_2009} reanalyzed the spectrum of the Cas~A CCO and conclude it is likely a NS with non-uniform surface temperature and low magnetic field. 

Cas A was observed by the Chandra Space telescope on May 05, 2012 for a total exposure time of 63.39 ks (ACIS-S; PI: Pavlov;\dataset[doi: 10.25574/cdc.484]{\doi{10.25574/cdc.484}}). A broadband spectrum grouped to a minimum of 15 counts per bin over the energy range 0.3-6.5 keV is used for this study. The spectrum of the CCO in Cas A is fit by a simple five-parameter absorbed blackbody for modeling the soft emission with an additional absorbed power-law component to model the hard emission generated by the putative hotspot. Due to the simplicity of the model used to fit its spectrum, Cas A is a \textit{representative} example useful for demonstrating the consistency between optimization algorithms since both local and global methods find the same best-fit solution over identical weakly-constrained search boundaries. The best fit parameters and their corresponding fit statistics found by $\tt{XSPEC}$ and $\tt{XFit}$ are shown in Table \ref{table:casa-table} along with the minimum and maximum search boundaries used for the search. 

\begin{figure}
\plotone{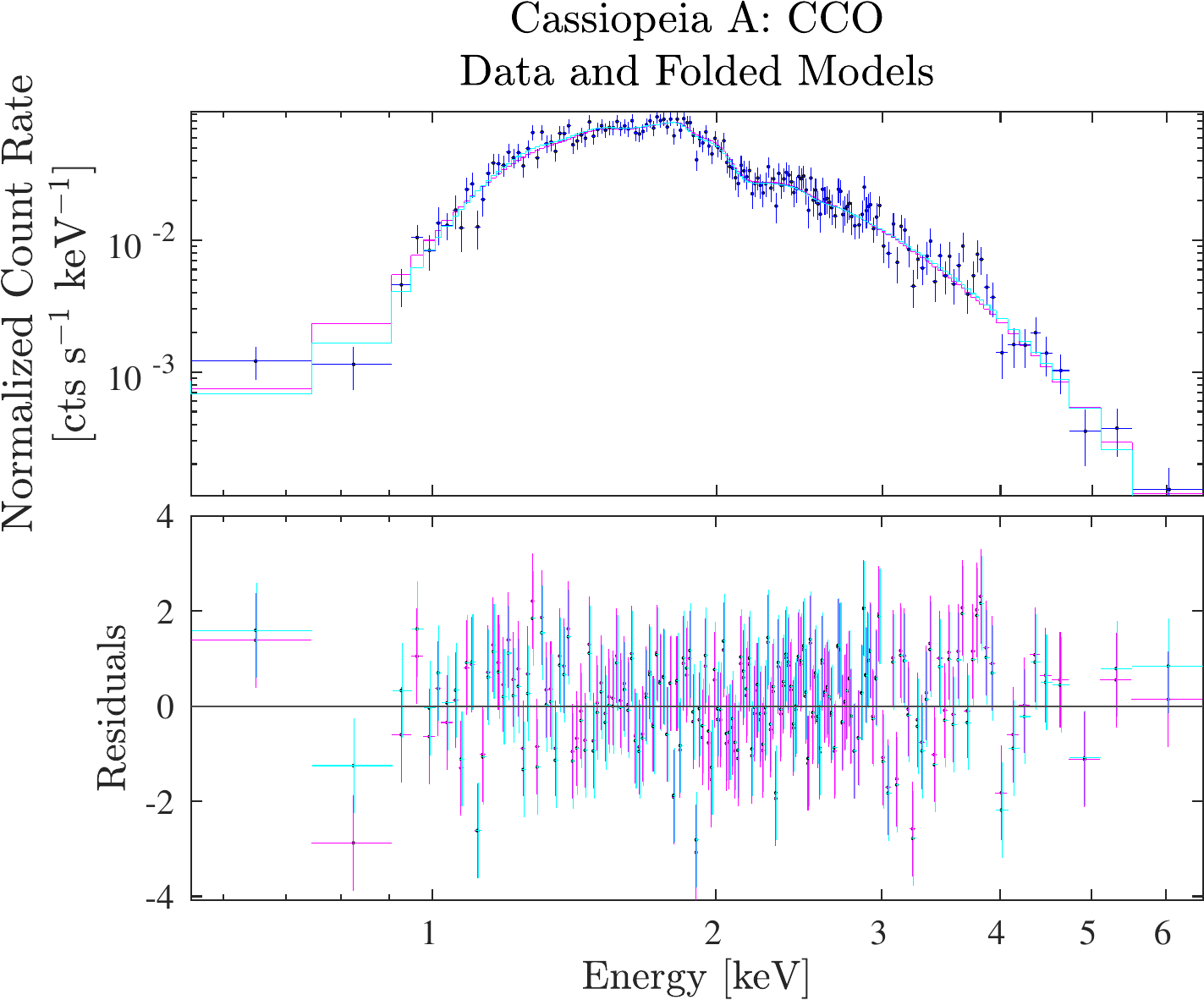}
\caption{Top panel: The spectral data and uncertainties of the CCO in Cassiopeia A is plotted logarithmically in dark blue crosses. The best-fit $\tt{XSPEC}$ model is plotted as a solid cyan line and the best-fit $\tt{XFit}$ model is plotted as a solid magenta line. Bottom panel: the residuals between the data and model for each optimizer. Residuals for each solution are plotted side-by-side at each energy bin to make nearly-identical pairs of data points both comparable and distinguishable}.\label{fig:casa-spectrum}
\end{figure}

Both weakly-constrained and unconstrained searches are useful in problems where one knows nothing about the expected range of parameter values containing the best-fit solution a priori. They are also useful when one wishes to map the topology of the parameter space in order to inform future searches, as well as in searching for degeneracies in best-fit solutions. Due to the simplicity of the model used in fitting CXOU J232327.9+584842, both the LMA and EO converge to the same minimum in the weakly-constrained search. The best fit model and residuals for the LMA (cyan) and EO (magenta) are plotted in Figure \ref{fig:casa-spectrum} showing strong agreement between the solutions found by each optimizer.

\begin{deluxetable*}{rcccc}
\tablecaption{CXOU J232327.9+584842 Best-fit Model Parameters \label{tab:casa-description}}
\tablehead{
\colhead{Parameter} & \colhead{Min. bound} & \colhead{Max. bound} & \colhead{XSPEC} & \colhead{XFit}
}
\startdata
$N_H$ $10^{22}$ $^a$  & 0.01 & 10       & $3.24_{-0.64}^{+0.89}$   & $3.24^{+0.55}_{-0.44}$   \\ \hline 
$kT$ $^b$ & 0.001                        & 100      & $0.41_{-0.02}^{+0.01}$   & $0.42^{+0.02}_{-0.02}$   \\
$N_{BB}$ $^c$ & $10^{-16}$                     & $10^{20}$ & $1.65_{-0.24}^{+0.86}\times 10^{-5}$ & $1.65^{+0.52}_{-0.26} \times10^{-5}$ \\ \hline
$\Gamma$ & 0.01                         & 10       & $5.56_{-1.00}^{+2.07}$   & $5.56^{+1.3}_{-0.73}$   \\
$N_P$ $^d$ & $10^{-16}$                     & $10^{20}$ & $5.97_{-4.17}^{+18.96}\times10^{-3}$ & $5.97^{+8.3}_{-3.1} \times10^{-3}$ \\ \hline
$\chi^2$       &                              &          & 159.57   & 159.57\\
DoF       &                              &          & 164      & 164      \\
$\#$ PHA bins  &                              &          & 169      & 169 \\ 
\enddata
\hspace{1mm} $^a$ $cm^{-2}$ \\
$^b$ $keV$ \\
$^c$ photons keV$^{-1}$ cm$^{-2}$ s$^{-1}$ \\
$^d$ photons keV$^{-1}$ cm$^{-2}$ s$^{-1}$ at 1 keV
\tablecomments{A comparison of the best-fit model solutions found by $\tt{XSPEC}$ and $\tt{XFit}$ for the  CCO in Cassiopeia A. The minimum and maximum search limits are given, as well as the parameters and associated fit statistic. The model is an absorbed blackbody and power-law. The column density of the absorption model is $N_H$, the temperature of the black-body component is given by $kT$ with normalization $N_{BB}$ and the power-law exponent is given by the photon index $\Gamma$ with normalization $N_P$. The fit statistic is $\chi^2$, number of degrees of freedom (DoF) over the stated number of pulse height PHA bins (pulse height amplitude: integrated charge per pixel from an event recorded in the detector).\label{table:casa-table}}
\end{deluxetable*}

\subsection{G41.1--0.3} \label{subsec:3c397}
\begin{figure*}
    \epsscale{0.97}    
    \plottwo{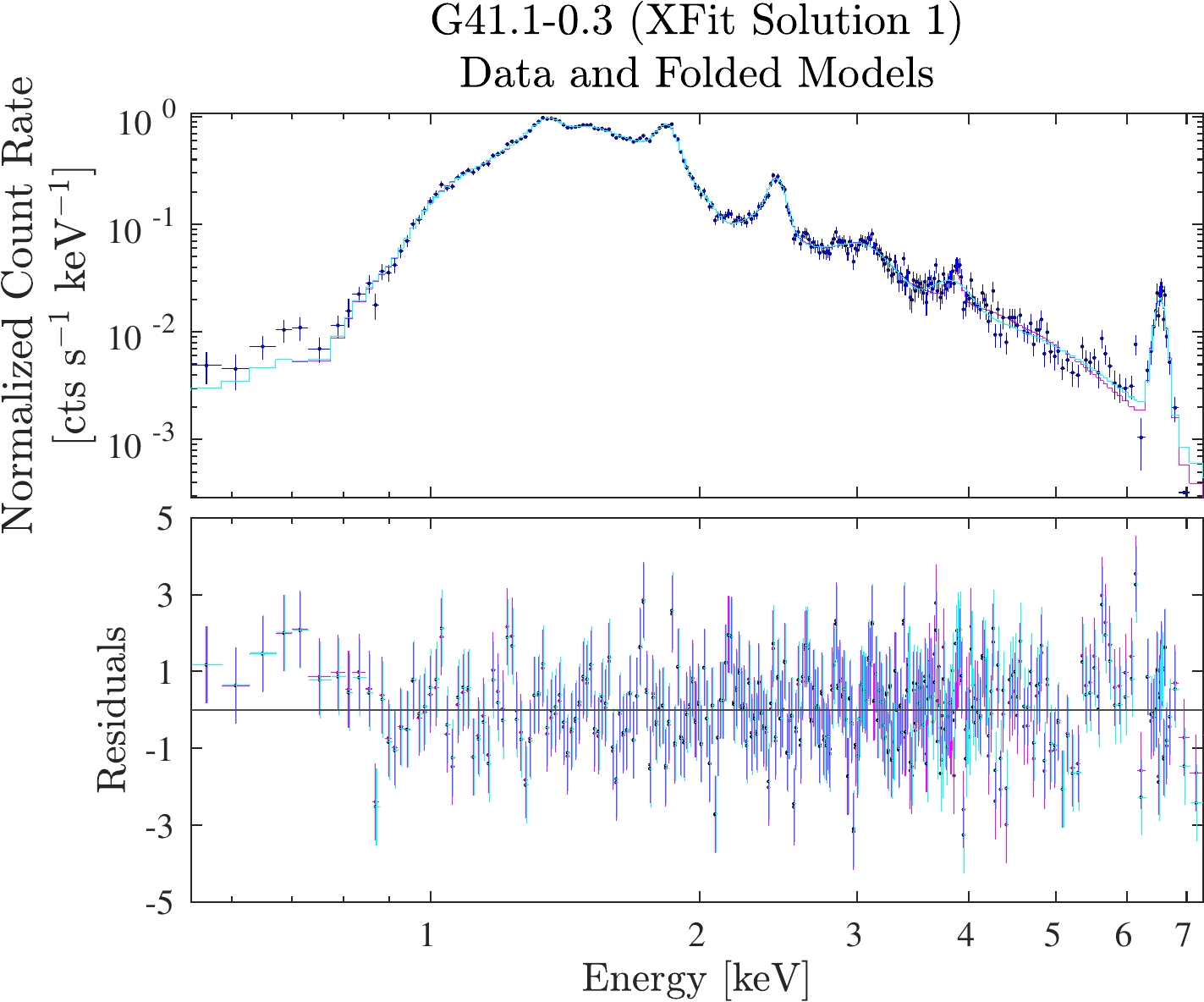}{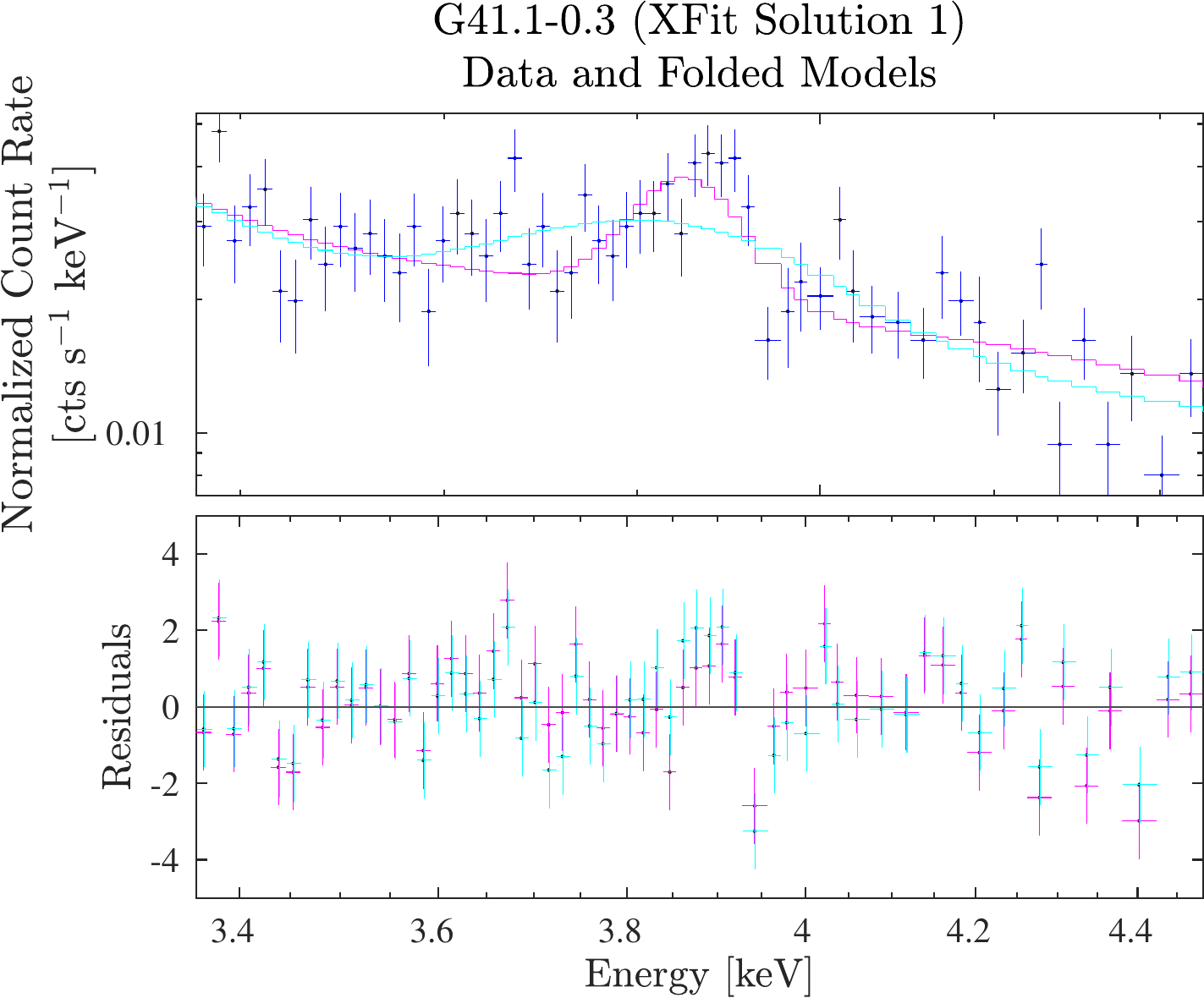}       
    \caption{\textit{Left}: Best-fit models and residuals fit to the spectrum of the western lobe of G41.1--0.3 are plotted logarithmically for Solution 1 ($\chi_r^2=1.238$). Observed count rates and their uncertainties are plotted as dark blue crosses. The best-fit $\tt{XSPEC}$ model is plotted as a solid cyan line and the best-fit $\tt{XFit}$ model is plotted as a solid magenta line. \textit{Right}: a zoomed-in view across a narrower range of energies highlighting degeneracies in spectral features of the best-fit Ca (He$\alpha$) amplitudes, line centroids and widths found by $\tt{XSPEC}$ and $\tt{XFit}$. Residuals for each solution are plotted side-by-side at each energy bin to make nearly-identical pairs of data points both comparable and distinguishable}.\label{fig:g41-sol1-spectrum}
\end{figure*}

\begin{figure}
    \epsscale{0.97}
    \plotone{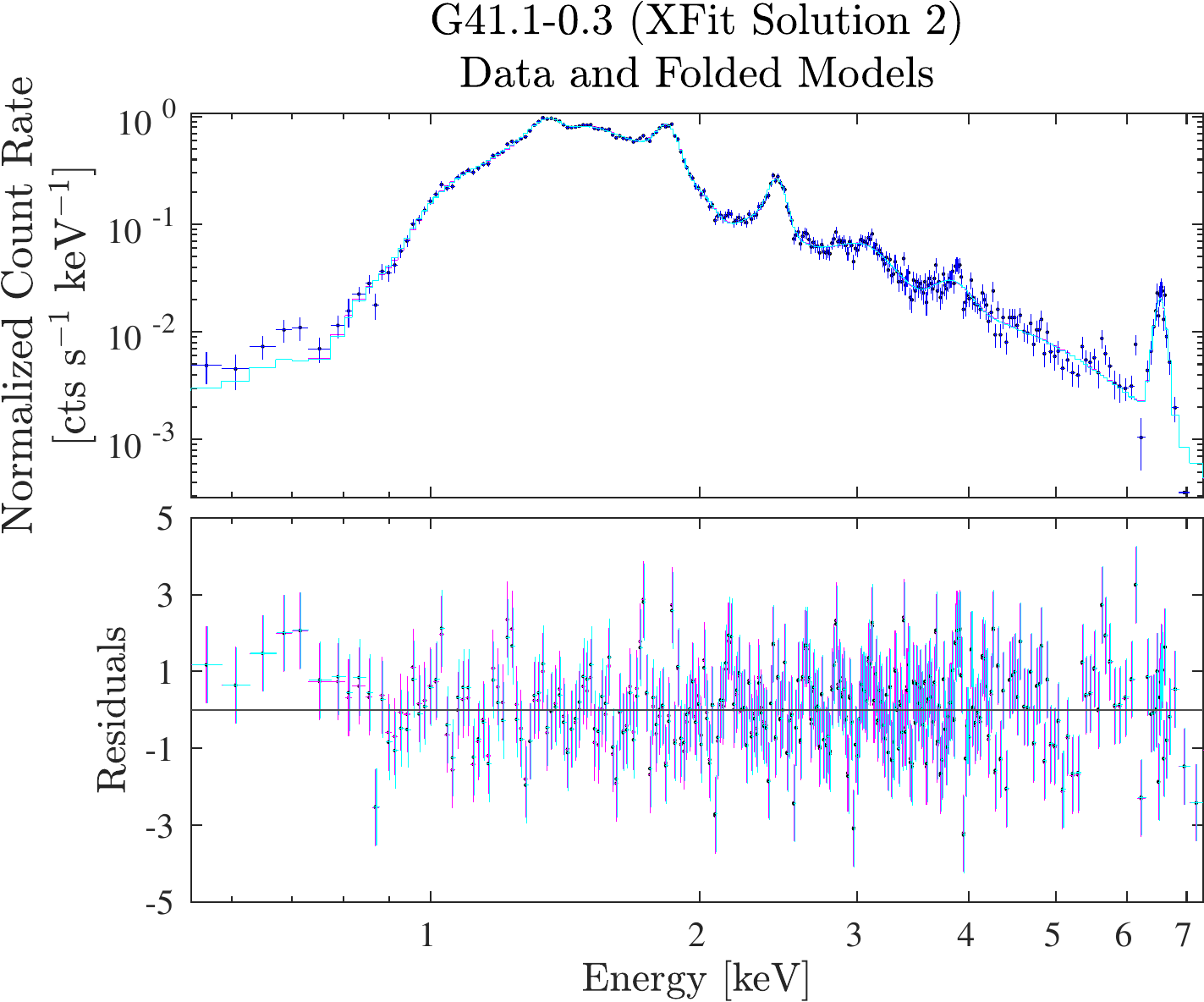}              
    \caption{Best-fit models and residuals fit to the spectrum of the western lobe of G41.1--0.3 are plotted logarithmically for Solution 2 ($\chi_r^2=1.225$). Observed count rates and their uncertainties are plotted as dark blue crosses. The best-fit $\tt{XSPEC}$ model is plotted as a solid cyan line and the best-fit $\tt{XFit}$ model is plotted as a solid magenta line. Residuals for each solution are plotted side-by-side at each energy bin to make nearly-identical pairs of data points both comparable and distinguishable}. \label{fig:g41-sol2-spectrum}
\end{figure}

The Galactic SNR 3C 397 (G41.1--0.3) is among the brightest of the Galactic radio SNRs due to its filled central X-ray emission, and is selected as a high-dimensional, representative example given that it is a uniquely bright and ejecta-dominated SNR whose supernova progenitor is being debated \citep{Yamaguchi2014}. A broadband study performed by \cite{Safi-Harb2000} used combined ROSAT, ASCA, and RXTE spectra to fit a two-component non-equilibrium ionization model to the soft- and hard-emission associated with 3C 397, finding a heavily absorbed spectrum dominated by thermal emission containing Mg, Si, S, Ar, and Fe emission lines. These findings were later confirmed by a spatially resolved Chandra observation \citep{Safi-Harb2005}. This composite SNR features a shell type remnant in the radio \citep{Green2004} and an X-ray bright central region. In this work we use broadband Chandra observations from the ACIS-S detector (66 ks exposure time taken on September 6, 2001; PI:Holt; \dataset[doi: 10.25574/cdc.484]{\doi{10.25574/cdc.484}}) with the spectrum grouped to a minimum of 20 counts per bin over the energy range 0.5-7.5 keV. The continuum and line emission of the western lobe of G41.1--0.3 is fit to a 29-parameter absorbed two-component Bremsstrahlung model with eight Gaussian peaks. This high-dimensional model provides a useful test of a more complicated model for the optimizers. Due to its high-dimensionality, the search-space hypervolume is constrained to a smaller range compared to those used to fit the Cas A CCO, including constraining the Bremsstrahlung model components into high- and low-temperature components. The search boundaries and best-fit parameter values found by both $\tt{XFit}$ and $\tt{XSPEC}$ for the western lobe model can be found in Table \ref{table:g41-table}. The data and best-fit models found by both optimizers are shown in Figures \ref{fig:g41-sol1-spectrum} and \ref{fig:g41-sol2-spectrum} with a close-up of the fit to the Ca emission line shown in the rightmost panels of Figure \ref{fig:g41-sol1-spectrum}.

\begin{deluxetable*}{rccccC}
\tablecaption{Western Lobe of G41.1--0.3 Best-fit Model Parameters \label{tab:g41-description}}
\tablehead{
\colhead{Parameter} & \colhead{Min. bound} & \colhead{Max. bound} & \colhead{$\tt{XSPEC}$ Solution 1} & \colhead{$\tt{XFit}$ Solution 1} & \colhead{$\tt{XFit}$ Solution 2}
}
\startdata
$N_H$ $^a$         & 4          & 8          & $4.39^{+0.84}_{-0.13}$ & $4.635^{+0.061}_{-0.057} $ & $5.27^{+0.18}_{-0.11} $ \\ \hline
$kT_1$ $^b$       & 0.01       & 1         & $0.17^{+0.01}_{-0.03}$ & $0.155^{+0.004}_{-0.002} $ & $0.135^{+0.005}_{-0.007} $ \\
$N_{BR1}$ $^c$     & $10^{-4}$  & $10^{5}$  & $3.40^{+52.2}_{-1.4}\times 10^{2}$ & $7.4^{+1.6}_{-1.9}\times 10^{2}$ & $5^{+5}_{-1.8}\times 10^{3}$ \\ \hline
$kT_2$ $^b$       & 1.1        & 5          & $3.3533^{+0.0001}_{-0.82}$ & $1.96^{+0.21}_{-0.17} $ & $3.24^{+1.2}_{-0.53} $ \\
$N_{BR2}$ $^c$    & $10^{-4}$  & $10^{5}$   & $1.1^{+0.64}\times 10^{-3}$ & $3.04^{+0.64}_{-0.55}\times 10^{-3}$ & $1.25^{+0.35}_{-0.41}\times 10^{-3}$ \\ \hline
$E_{1}$ $^b$      & 0.01       & 1.1        & $0.705^{+0.002}_{-0.15}$ & $0.693^{+0.028}_{-0.14} $ & $0.522^{+0.086}_{-0.034} $ \\
$\sigma_{1}$ $^b$  & $10^{-4}$  & 1         & $0.12^{+0.02}$ & $0.121^{+0.025}_{-0.006} $ & $0.142^{+0.007}_{-0.014} $ \\
$N_{G1}$  $^d$    & $10^{-6}$  & $10^{4}$   & $59^{+2324}_{-21}$ & $1.46^{+6.2}_{-0.4}\times 10^{2}$ & $9.52^{+0.48}_{-6.2}\times 10^{3}$ \\ \hline
$E_{2}$ $^b$      & 1.2        & 1.4        & $1.3351^{+0.003}_{-0.0002}$ & $1.336^{+0.008}_{-0.004}$ & $1.335^{+0.001}_{-0.003} $ \\
$\sigma_{2}$ $^b$  & $10^{-4}$  & 1         & $3.3^{+20}_{-2}\times 10^{-4}$ & $2.81^{+11}_{-2.7}\times 10^{-3}$ & $0.0232^{+1.4}_{-0.013}\times 10^{-2}$ \\
$N_{G2}$ $^d$     & $10^{-6}$  & $10^{4}$   & $8.5^{+15}_{-0.8}\times 10^{-3}$ & $1.163^{+0.1}_{-0.1}\times 10^{-2}$ & $2.6^{+0.67}_{-0.31}\times 10^{-2}$ \\ \hline
$E_{3}$ $^b$      & 1.8        & 1.95       & $1.830^{+0.0005}_{-0.002}$ & $1.830^{+0.002}_{-0.002} $ & $1.829^{+0.002}_{-0.004} $ \\
$\sigma_{3}$ $^b$  & $10^{-4}$  & 1         & $2.8^{+0.5}_{-0.8}\times 10^{-2}$ & $2.70^{+0.38}_{-0.43}\times 10^{-2}$ & $3.33^{+0.49}_{-0.32}\times 10^{-2}$ \\
$N_{G3}$ $^d$     & $10^{-6}$  & $10^{4}$   & $9.9^{+7.6}_{-1.3}\times 10^{-4}$ & $1.122^{+0.043}_{-0.069}\times 10^{-3}$ & $1.71^{+0.25}_{-0.14}\times 10^{-3}$ \\ \hline
$E_{4}$ $^b$      & 2.2        & 2.35       & $2.27^{+0.03}_{-0.09}$ & $2.268^{+0.019}_{-0.011} $ & $2.181^{+0.02}_{-0.077} $ \\
$\sigma_{4}$ $^b$  & $10^{-4}$  & 1         & $0.209^{+0.06}_{-0.007}$ & $0.217^{+0.017}_{-0.018} $ & $0.267^{+0.061}_{-0.03} $ \\
$N_{G4}$ $^d$     & $10^{-6}$  & $10^{4}$   & $5.2^{+5.4}_{-0.8}\times 10^{-4}$ & $5.16^{+0.48}_{-0.65}\times 10^{-4}$ & $1.05^{+0.38}_{-0.14}\times 10^{-3}$ \\ \hline
$E_{5}$ $^b$      & 2.35       & 2.95       & $2.4303^{+0.0005}_{-0.0003}$ & $2.430^{+0.004}_{-0.001} $ & $2.430^{+0.004}_{-0.001} $ \\
$\sigma_{5}$ $^b$  & $10^{-4}$  & 1         & $1.1^{+5.4}_{-0.9}\times 10^{-3}$ & $0.09156^{+1.4}_{-0.082}\times 10^{-2}$ & $0.0145^{+1.4}_{-0.0045}\times 10^{-2}$ \\
$N_{G5}$ $^d$     & $10^{-6}$  & $10^{4}$   & $1.8^{+0.4}_{-0.1}\times 10^{-4}$ & $1.961^{+0.091}_{-0.16}\times 10^{-4}$ & $2.32^{+0.24}_{-0.11}\times 10^{-4}$ \\ \hline
$E_{6}$ $^b$      & 2.35       & 3.11      & $2.9515^{+0.06}_{-0.0001}$ & $2.977^{+0.024}_{-0.018} $ & $2.979^{+0.044}_{-0.045} $ \\
$\sigma_{6}$ $^b$  & $10^{-4}$  & 1        & $0.23_{-0.05}$ & $0.191^{+0.019}_{-0.022} $ & $0.217^{+0.036}_{-0.029} $ \\
$N_{G6}$ $^d$     & $10^{-6}$  & $10^{4}$  & $1.623^{+0.004}_{-0.4}\times 10^{-4}$ & $1.083^{+0.088}_{-0.14}\times 10^{-4}$ & $1.67^{+0.41}_{-0.28}\times 10^{-4}$ \\ \hline
$E_{7}$ $^b$      & 3.75       & 3.95      & $3.8062_{-0.01}$ & $3.846^{+0.023}_{-0.015} $ & $3.793^{+0.039}_{-0.039} $ \\
$\sigma_{7}$ $^b$  & $10^{-4}$  & 1        & $0.17^{+0.01}_{-0.01}$ & $8.07^{+39}_{-8}\times 10^{-3}$ & $0.181^{+0.039}_{-0.034} $ \\
$N_{G7}$ $^d$     & $10^{-6}$  & $10^{4}$  & $2.5^{+0.4}_{-0.3}\times 10^{-5}$ & $9.9^{+1.7}_{-1.5}\times 10^{-6}$ & $2.87^{+0.77}_{-0.42}\times 10^{-5}$ \\ \hline
$E_{8}$ $^b$      & 6.53       & 6.56      & $6.546^{+0.004}_{-0.007}$ & $6.547^{+0.013}_{-0.013} $ & $6.546^{+0.012}_{-0.013} $ \\
$\sigma_{8}$ $^b$  & $10^{-4}$  & 1        & $6.35^{+0.02}_{-0.3}\times 10^{-2}$ & $7.1^{+1}_{-1.9}\times 10^{-2}$ & $6.3^{+1.3}_{-1.5}\times 10^{-2}$ \\
$N_{G8}$ $^d$     & $10^{-6}$  & $10^{4}$  & $4.10^{+0.1}_{-0.01}\times 10 ^{-5}$ & $4.32^{+0.46}_{-0.48}\times 10^{-5}$ & $4.14^{+0.37}_{-0.4}\times 10^{-5}$ \\ \hline
$\chi^2$       &                              &          & 323.48   & 326.81  & 323.51 \\
\# DoF       &                              &          & 264      & 264    & 264  \\
\# PHA bins  &                              &          & 293      & 293       & 293  \\ \hline
\enddata
\hspace{1mm} $^a$ $cm^{-2}$ \\
$^b$ $keV$ \\
$^c$ photons keV$^{-1}$ cm$^{-2}$ s$^{-1}$ \\
$^d$ photons keV$^{-1}$ cm$^{-2}$ s$^{-1}$ at 1 keV
\tablecomments{A comparison of $\tt{XSPEC}$ and $\tt{XFit}$ best-fit model solutions for the western lobe of G41.1--0.3. The minimum and maximum search limits are given as well as the parameters and associated fit statistic. The model is a two-component bremsstrahlung model with eight Gaussian lines. The column density of the absorption model is $N_H$, the temperature of each bremsstrahlung component is given by $kT_{n}$ with normalization $N_{BRn}$ with n denoting the component index. The centroid energy of each Gaussian line component is given by $E_{n}$, and the line width and normalization are given by $\sigma_{n}$ and $N_{Gn}$ respectively. The fit statistic is $\chi^2$, number of degrees of freedom (DoF) over the stated number of PHA bins.\label{table:g41-table}}
\end{deluxetable*}

\section{Results\label{sec:results}}

Due to their differing approaches to optimization, comparisons between local and global optimization algorithms must be performed using measurable quantities that are common to both types of algorithm, such as the number of evaluation calls to the objective function. This quantity is used to determine the performance of the search at each step and inform the direction of future steps, with a `step', referring to a change in the coordinate of a search-agent in the parameter space. Depending on both the optimization algorithm, and the function being optimized, numerous objective function evaluations may be required to determine the next-best coordinate for a search-agent as the optimization continues. Thus, the number of objective function evaluations is always equal to, or greater than the number of steps in an optimization.

As discussed in Section \ref{sec:intro}, local optimizers such as the LMA use a single-agent gradient-based search approach that stores information about the position, gradient, and fitness value only at the current and immediately neighboring time steps, and disposes of any information obtained at previous steps. In contrast, EOs collect and store any information that contributes to a decrease in the objective function throughout the entire optimization and exchanges this information usefully among its search agents. For these reasons, we are interested in comparisons between the number of degenerate minima found by each optimizer, and the frequency at which such solutions are found for a given number of objective function evaluations. Following this argument, the EO and LM algorithms are allotted an equal number of calls to the objective function for evaluation. The initial population of parameter sets used in the EO search is randomly sampled from a distribution with user-defined limits for each parameter, and boundary values that are given in Tables \ref{table:casa-table} and \ref{table:g41-table}. $\tt{XFit}$ then performs a search, converges on one or more solutions, and returns the total number of objective function evaluations N in a run. An initial starting point for the LM search is then randomly sampled from a uniform distribution with the same boundaries as was used by $\tt{XFit}$. 

The single-step stopping criterion for the LMA works by signaling a stop as soon as the improvement in the fitness falls below a user-defined absolute threshold value termed the `critical delta', $\Delta_\text{crit}$. The value for the critical delta $\Delta_{\text{crit}}=0.001$ was selected to be of the same order as the absolute difference in the best objective function values found by $\tt{XFit}$ at the penultimate and ultimate generations of its optimization, ensuring that both optimizers are capable of distinguishing between solutions at the same level of precision in the convex basin. An additional stopping criterion for the LMA is employed called the `fit delta' $\Delta_\text{fit}$, used to prevent wasting evaluations in regions where parameter values aren't changing significantly even if $\Delta_\text{crit}$ has not yet been triggered. $\Delta_\text{fit}$ is related to the relative difference in a parameter's value between successive steps. When this relative difference falls below a user-defined threshold for all model parameters, $\Delta_\text{fit}$ is triggered. If either of these stopping criteria are triggered, the LMA is signaled to halt, and the current solution is returned. Trajectories taken by the LMA through the parameter space are deterministic, since for identical initial values for the damping parameter $\lambda$, step size $h$, $\Delta_\text{crit}$, and $\Delta_\text{fit}$ at the start of the search, any solution found that triggers stopping solely depends on the initial starting point of the search. We use a default value of $\Delta_\text{fit}=0.001$. The parameter deltas used for calculating the finite-difference derivatives of the Jacobian matrix are set to their default values. Once the stopping criterion for an LMA search is reached, statistics about the run are collected, including a tally of the number of objective function evaluations called during the run. If the total number of objective function evaluations called at the end of the current LMA run is less than the N calls made by the EO, another LMA search is initiated with a newly sampled starting point and this process is continued until N total objective function evaluations have been called by the LMA. The uncertainties in the best-fit model parameters found by each optimizer listed in Tables \ref{table:casa-table} and \ref{table:g41-table} are determined using the entire set of solutions found within the 90\% confidence interval of the minimum $\chi_r^2$ solution.

All experiments were performed on a workstation equipped with an AMD Ryzen Threadripper PRO 5995WX CPU (64 cores, 2.7 GHz base/4.5 GHz boost) and 256 GB of RAM \citepalias{passmark5995wx}, using MATLAB with the Parallel Computing Toolbox and 64 workers for parallel evaluation of the population. We benchmarked the scaling of $\tt{XFit}$’s runtime with the number of free parameters $N$ by varying $N$ from 1 to the full $N_{max}=29$. Over this range, the mean wall-clock time per fitness evaluation remains in the $\sim 1$-$2$ ms regime and is well described by an approximately linear model with a fixed overhead of $\sim0.7$ ms due to the MEX API loading time, and an additional $\sim 30$ $\mu$s per free parameter. For the full 29-parameter model, a complete optimization run with our standard settings requires a total wall-clock time of $\sim 7$ hours on this system.

\subsection{CXOU J232327.9+584842}

\begin{figure*}
        \epsscale{0.97}
        \plottwo{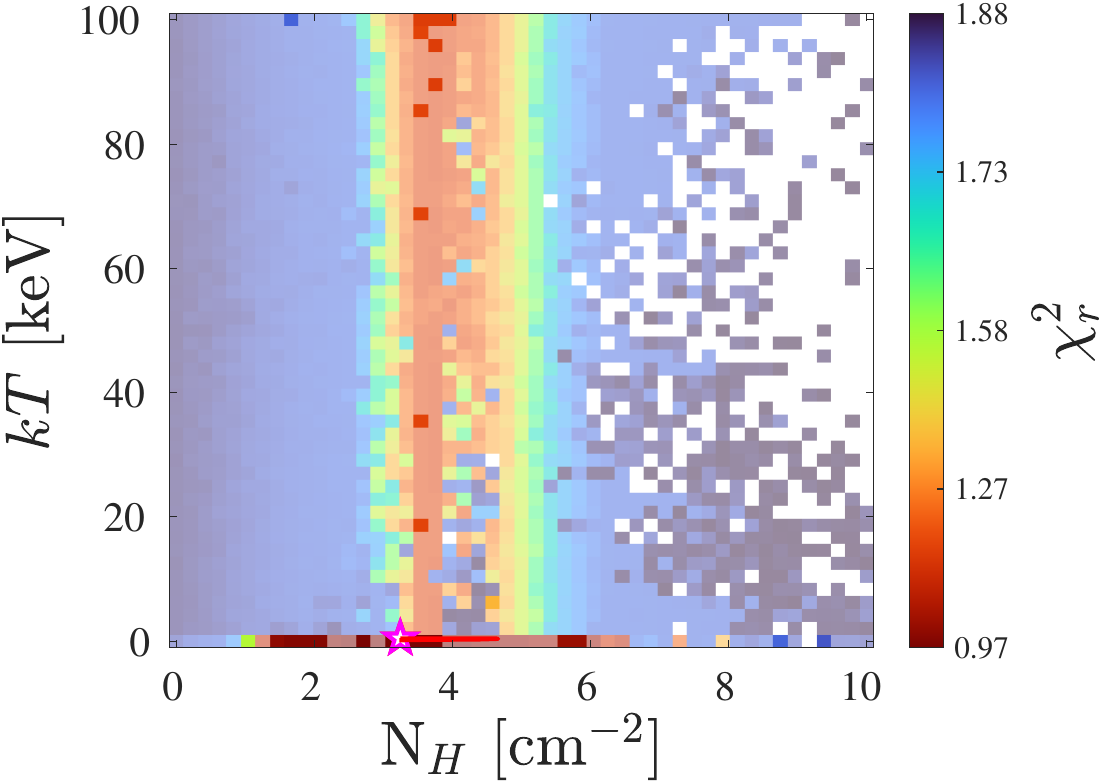}{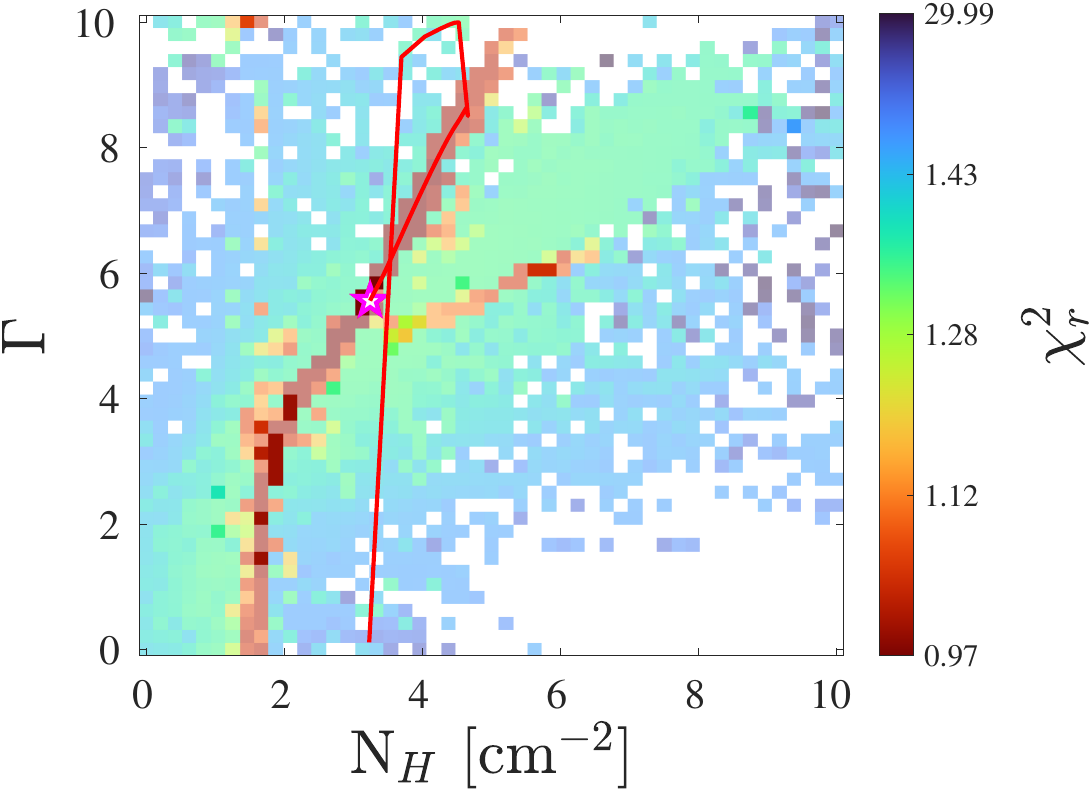}          
    \caption{Parameter space projections of the column density $N_H$, blackbody temperature $kT$, and photon index $\Gamma$ for the absorbed thermal model used to fit the spectrum of CXOU J232327.9+584842, visualizing  the entire space of points covered by the 5365 independent searches conducted by the LMA. The data is spatially binned to include only the best-fit $\chi^2_r$ solution found in each bin. Coordinates with high color saturation coincide with local minima in the parameter space at which the LMA converged. The small proportion of local minima found by the local optimizer is representative of the `simplicity' of the model. $\tt{XFit}$'s best found solution is marked with a magenta star and $\tt{XSPEC}$'s best found solution is marked with a red star which is covered by $\tt{XFit}$'s solution since both optimizers converge to the same global best value. A trailing red line shows the path taken by the local optimizer in its search to find the best solution starting from the initial optimization step, and illustrates the strong tendency of the LMA to move along narrow `canyons' with large negative gradients.\label{fig:casa-xspec}}
\end{figure*}

In fitting the spectrum of CXOU J232327.9+584842, $\tt{XFit}$ was given four populations of 100 search agents, finding the global best solution and uncertainties (90\% confidence) shown in Table \ref{table:casa-table} after $\text{N}_{\text{CasA}}=582592$ objective function evaluations. The LMA completed 5365 independent local runs before reaching the maximum number of fitness calls returned by $\tt{XFit}$, finding a global best solution that is in strong agreement with the solution found by $\tt{XFit}$. Figure \ref{fig:casa-xspec} displays a parameter space map of solutions discovered by the LMA across N evaluation. To ensure the map only contains solution arrived at by the LMA using gradient information, the initial starting points of each search are excluded. Highly colour-saturated bins coincide with local minima, the final coordinate of an LMA trajectory at the moment the stopping criterion was triggered. Each coordinate bin is colour-coded based on the best $\chi^2$ value attained within that bin. The clustering of solutions around local minima observed in the maps in Figure \ref{fig:casa-xspec} illustrates the local optimizer’s sensitivity to the initial starting conditions of the run. We find very few regions corresponding to the final positions of the search indicating that this is a simple problem with few local minima. This simplicity is further confirmed by the matching solutions found by both local and global optimizers. If one knows very little about a given model or would like to avoid introducing bias into the results of a run, these maps can provide insights into suitable search boundaries for an optimizer, resulting in a potential reduction in required resources and computation time. 

Uncertainties in model parameters are obtained using the set of all solutions found by each optimizer with a fit statistic $\chi^2$ within the 3$\sigma$ (99.7\%) confidence interval, which we refer to as the “optimal set”. In general, the optimal set can possess complicated morphologies or even be spread out over multiple, disconnected regions of parameter space. Confidence maps are important for determining the precision with which we can say our solution best reflects the true values of the physical properties of the source as well as for mapping the behaviour of an optimizer in the fitness landscape surrounding local minima. In problems where multiple degenerate solutions exist, confidence mapping also allows one to distinguish between solutions with fitness values within the 3$\sigma$ range of the global best.

Mapping confidence regions in $\tt{XSPEC}$ is typically achieved by creating an M-dimensional grid of linearly spaced values around a previously found solution for the number of “interesting” parameters requiring confidence estimates. The parameters of interest are held at constant values corresponding to their position in the grid and the remaining ``uninteresting'', free parameters are given initial values corresponding to the latest best fit solution. A local fit is then performed at each point in the grid by varying the “uninteresting” parameters until a new best-fit solution is achieved and an interpolation over all points generates contours at the 1$\sigma$, 2$\sigma$ and 3$\sigma$ levels above the best-fit solution. Due to its reliance on local optimization, this procedure inevitably runs into the same caveats discussed in Section \ref{sec:intro}. For a parameter space with many local minima, the question also arises as to how reliably the contours generated using this method map the space around solutions. $\tt{XFit}$ simply maps confidence regions by returning all individual solutions that are members of the optimal set found during a run. This contrasts with $\tt{XSPEC}$s approach described above, which aims only to return a single best-known solution. Thus, $\tt{XFit}$ thoroughly and automatically discovers and maps the structure of the 1$\sigma$, 2$\sigma$ and 3$\sigma$ confidence interval projections while simultaneously searching for the best fit. This also allows $\tt{XFit}$ to map degenerate solutions and evaluate the uncertainties in their parameter values simultaneously.

\begin{figure*}
        \epsscale{0.5}
        \plotone{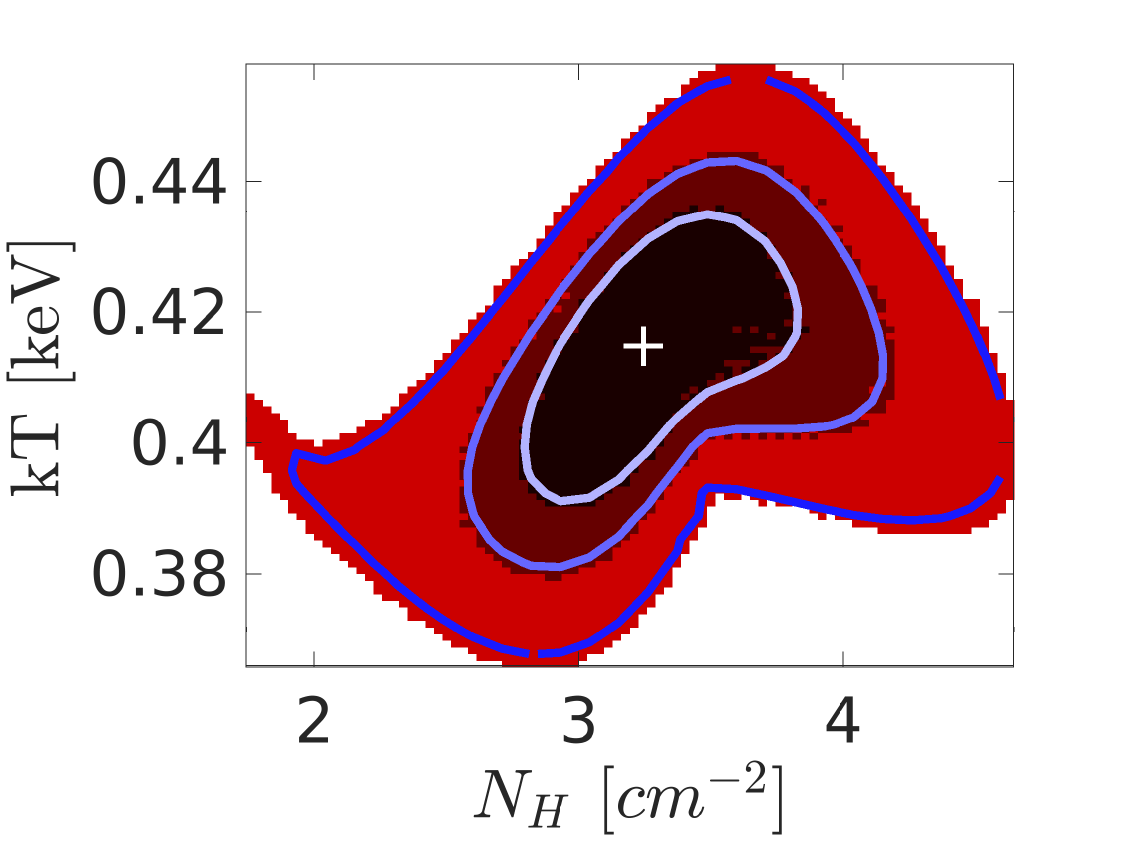}
        \epsscale{1}
        \plottwo{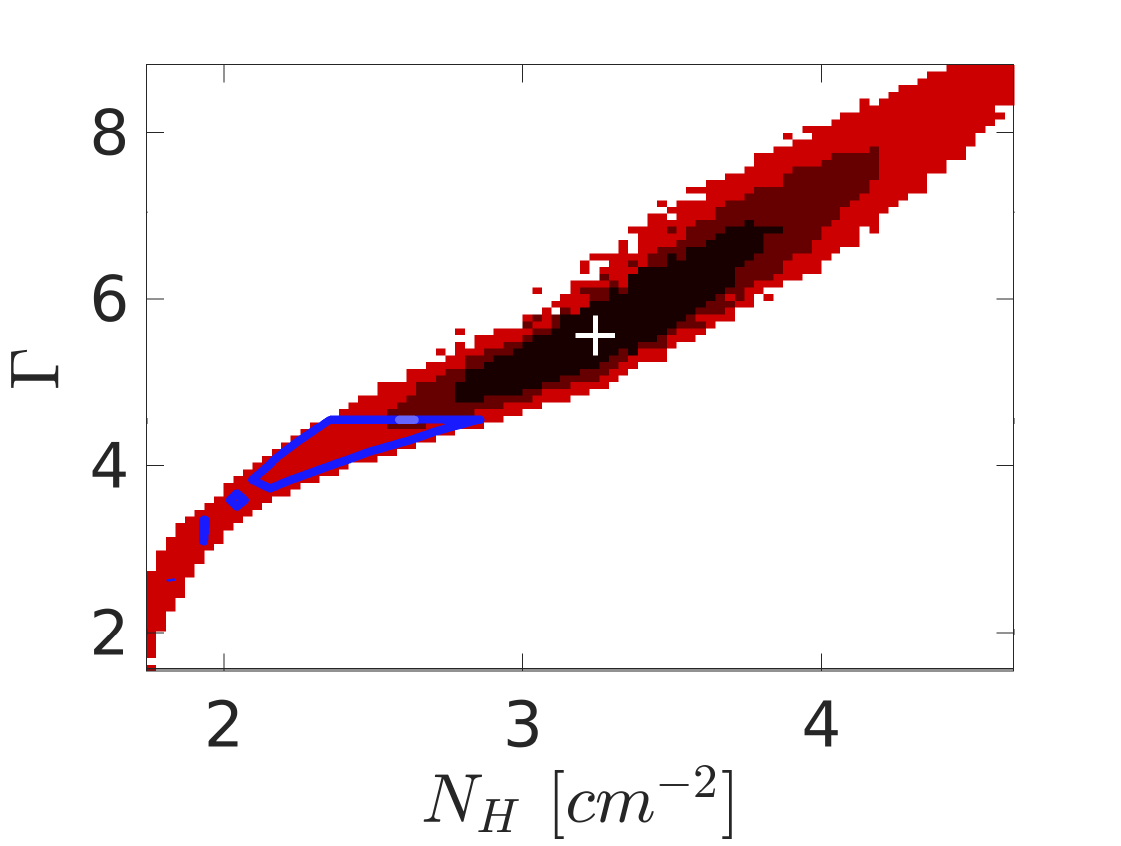}{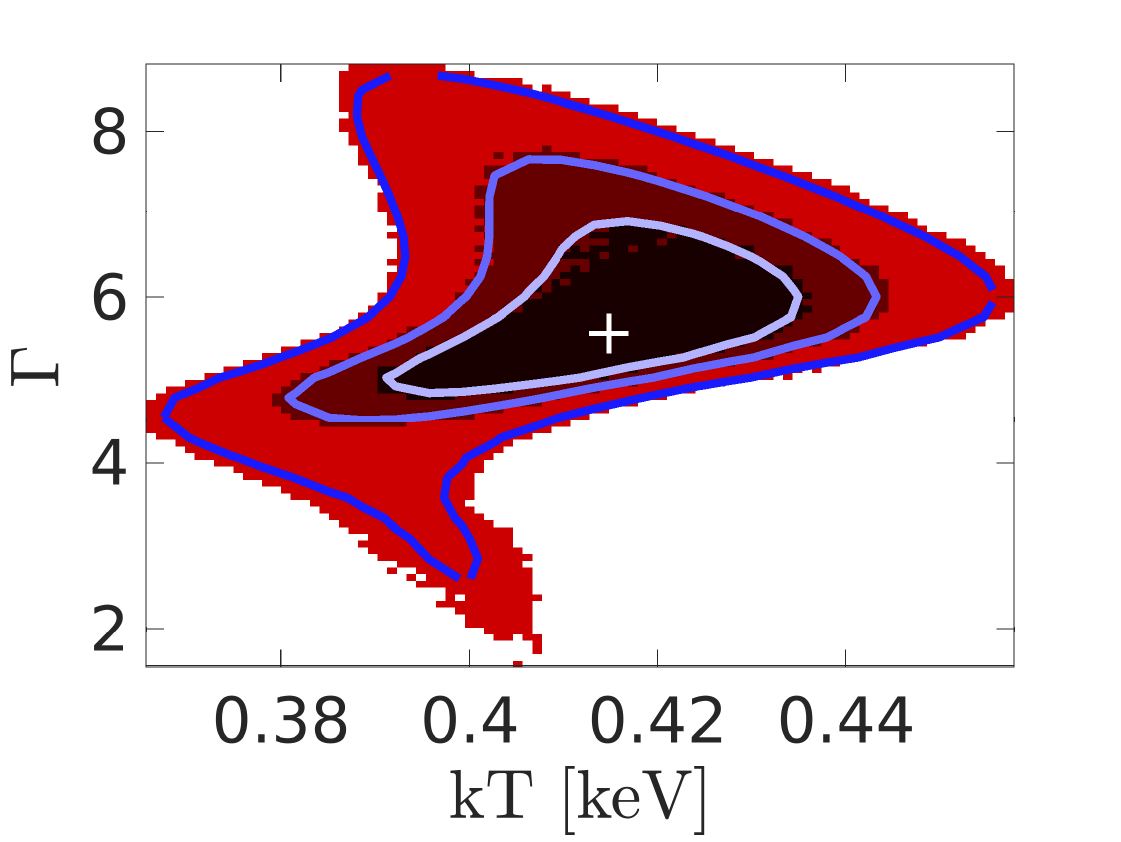}              
    \caption{A selection of parameter-space projections between the column density $N_H$, blackbody temperature kT, and photon index $\Gamma$ for the absorbed thermal model used to fit the spectrum of CXOU J232327.9+584842. The solutions discovered by $\tt{XFit}$ are plotted as $1\sigma$ (black), $2\sigma$ (dark red) and $3\sigma$ (red) areas. 480864 solutions are plotted and spatially binned and the best-fit solution found by $\tt{XFit}$ is marked by a white cross. The contours generated by $\tt{XSPEC}$ are plotted as $1\sigma$ (light blue), $2\sigma$ (blue) and $3\sigma$ (dark blue) lines. For this relatively simple model, the results of the two fitting methods coincide across the majority of parameters. However, the local optimizer appears to run into some difficulty mapping the space for the $N_H$ --- $\Gamma$ cross-section.\label{fig:casa-conf-grid}}
\end{figure*}

Figure \ref{fig:casa-conf-grid} shows the confidence interval maps generated by $\tt{XFit}$ for the CCO in Cas A with the 1$\sigma$, 2$\sigma$, and 3$\sigma$ intervals represented by red bins of varying darkness. Confidence contours surrounding the LMA's best-fit solution were then generated using $\tt{XSPEC}$'s conventional interpolation method and are plotted as solid blue and purple contour lines. The LMA and EO predict identical uncertainties in their parameters for the majority of the confidences and combinations of parameters when using this conventional approach, demonstrating that both optimizers observe the same solution and surrounding parameter space for CXOU J232327.9+584842. However, there is a notable disagreement between the optimizers for both confidence maps containing the power law index $\Gamma$. The 3$\sigma$ contour (dark-blue) generated by the local optimizer encounters difficulty in tracing the contours generated by the EO. A possible explanation for this discrepancy is the LMA becoming stuck in a local minimum or continuing the search along a narrow and steeply descending region in the parameter space, taking the search outside the allowed range of values defined in the grid. 

When performing many local searches in the manner described at the beginning of this section, it is not possible to predict the number of objective function evaluations that will be required to determine parameter uncertainties for the best-fit solution found by the LMA. Additionally, since $\tt{XSPEC}$'s interpolation method is not a built-in feature of the LMA, it is only used to confirm consistency between optimization methods for this simple model as shown in Figure \ref{fig:casa-conf-grid}, and is excluded from the remaining analysis. Instead, the uncertainties listed in Table \ref{table:casa-table} are determined using the set of solutions found by each optimizers after N objective function evaluations.

\subsection{G41.1--0.3}

The 29-parameter model used to fit the spectrum of G41.1--0.3 presents a more challenging case for both optimizers even with more narrowly constrained search limits. To allow for comparisons with previously published values \citep{Safi-Harb2005}, we here select the western lobe of this SNR. $\tt{XFit}$ performed its optimization using 6 populations of 250 individuals each, over a total of $\text{N}_{\text{EO}}=20285430$ objective function calls, and finds two unique solutions, indicating degeneracy in the parameter space of the model. The LMA completed 45297 independent local runs given the same number of objective function evaluations as $\tt{XFit}$, and finds a best-fit solution of $\min{\chi^2_{r}}=1.225$; a column density $N_H$ of $4.39$ cm$^{-2}$; temperatures of 0.17 keV and 3.4 keV for the first and second Bremsstrahlung components $kT_1$ and $kT_2$, respectively; values of 3.8 keV and 0.17 keV for centroid energy E$_7$ and line width $\sigma_7$ components of the seventh Gaussian line respectively, corresponding to the Ca He$\alpha$ emission line; and 0.24 keV for line width component of the sixth Gaussian line, corresponding to the Ar He$\alpha$ emission line. $\tt{XFit}$ also finds this solution ($\chi_r^{2}=1.225$), plus an additional, degenerate solution ($\chi^2_r=1.238$) within the $3\sigma$ uncertainty of $\min\chi^2_r$, with values of 4.635 $\text{cm}^{-2}$, 0.155 keV, 1.96 keV, 3.846 keV, $8.07\times10^{-3}$ keV, and 0.191 keV for these parameters, respectively. $\tt{XFit}$ also finds degeneracy in the emission line normalizations for all but the Fe line component $N_{G8}$. These findings have implications in SNR progenitor diagnostics, owing to their use in calculating emission line flux \citep{2020Martinez}. The lack of degeneracy in the Fe K$\alpha$ line is likely influenced by its high signal-to-noise in the spectrum, owing to its high abundance among the high-Z elements and its sensitivity to the hottest and most energetic astrophysical processes. Its insensitivity to foreground extinction also makes Fe ions among the strongest emitters in X-rays \citep{2020H}. This inverse correlation between the S/N of a model component and the expected degree of degeneracy in the solution space reinforces the role of Fe as a robust diagnostic element, particularly in the era of ultra-high-resolution X-ray microcalorimeter spectroscopy.

\begin{figure*}   
        \plotone{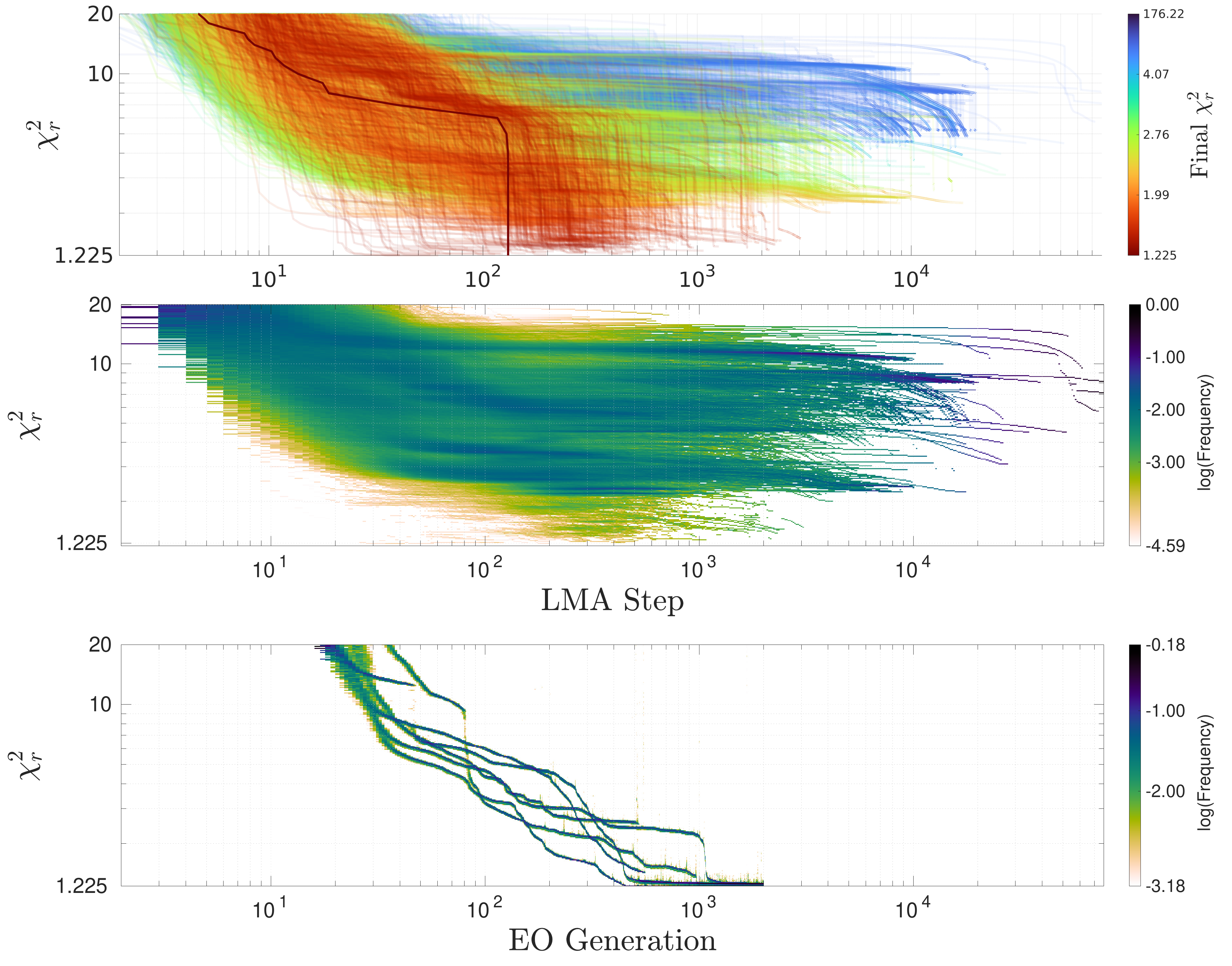}
    \caption{\textit{Top panel}: $\chi_r^2$ trajectories per optimization step of the 45297 LMA attempts performed by the local optimizer. Each path is monotonically color-coded based on the best local minimum fitness value found for each attempt. There is an inverse correlation between the number of steps taken by the optimizer and the best $\chi_r^2$ found for a given run, indicating the LMAs strong preference for steeply decreasing regions of the parameter space. The dark red line illustrates the path taken by the best solution found by the LMA. \textit{Middle} and \textit{Bottom} Panels: $\chi_r^2$ densities at each step in the LMA and EO optimizations respectively. Each coordinate in the grid is color-coded based on the frequency of binned $\chi_r^2$ values found within the step bin. \label{fig:G41-chi-per-gen}}
\end{figure*}

The top panel of Figure \ref{fig:G41-chi-per-gen} shows the best fitness value at each local optimization step for every search attempt made by the LMA. Searh trajectories are monotonically colour-coded based on the best-fit value found at the final step in the search. This plot demonstrates local and global optimizer behaviour that is in alignment with the ideas explored in Section \ref{sec:intro}. The LMA appears to tend towards a larger number of steps for large (worse) final fit statistics. Such searches represent the majority of the solutions found by the LMA and are likely becoming trapped in regions that span large volumes of the parameter space at high $\chi_r^2$ values, and subsequently declining rapidly towards unreliable fits, while also failing to trigger the stopping criterion close to the local minimum. For the small number of solutions found by the LMA that tend toward good $\chi^2$ values, the more quickly the run converges, the better the final fit statistic is, implying that success with the LMA depends strongly on choosing a fortuitous starting point and highlighting the risk of introducing bias into the final solution based on how this starting point is selected. 

An EO encodes information about the features of the parameter space that lead to improvements in the objective function and shares this encoded information usefully among individuals in a population at each generation, which is then passed on to offspring that populate future generations in the search. The various evolutionary operators discussed in Section \ref{sec:xfit} make it difficult to identify unique trajectories corresponding to an ``individual'' search agent's progress towards a final solution in the same way that can be done for the LMA's independent searches. Although such an analysis is currently outside the scope of the current work, it is instructive to consider representations that are common to both local and global optimizations. For example, distributions of the densities of solutions found by an optimizer each time $\textbf{p}$ is updated to a new coordinate in the parameter space, represented by a ``step'' and ``generation'' in the LMA and EO respectively, and can be used for more direct comparisons. The middle panel of Figure \ref{fig:G41-chi-per-gen} shows the distributions of $\chi_r^2$ values found by the LMA at each optimization step for fitness values $\chi_r^2 \leq 20$. A histogram is then used to color-code the distribution at each step as the fractional frequency of the total number of solutions found within a $\chi_r^2$ bin at that step. Since this visualization makes no use of tracing the line connecting any two subsequent steps made by a search agent, it can also be used to identically represent the search-space as seen by the EO, shown in the bottom panel of Figure \ref{fig:G41-chi-per-gen}. The EO uses significantly less computations in regions of high $\chi_r^2$ when compared with the LMA, as the trajectories taken by the EO show more compactness across values of $\chi^2_r$ than the LMA, illustrated by the six trajectories in the bottom panel of Figure \ref{fig:G41-chi-per-gen}. The EO trajectories show a strong amount overlap at early and late times in the run, marking epochs of `exploitation', wherein the optimizer favours fast convergence to small objective function values. Epochs of `exploration' are marked by the diverging of population trajectories, where the EO spreads itself across distinct regions of the parameter space in favour of collecting useful information about its topology to aid in future convergence, and can be seen to occur between roughly 30 and 1000 generations. Various points in this step range are marked by sudden, discontinuous drops in $\chi_r^2$ across populations. Discontinuities leading to convergence with another population is likely due to $\tt{XFit}$'s built-in `immigration' operator, that probabilistically selects individuals from one population to join another. Other discontinuous trajectories are preceded by sharp spikes in $\chi_r^2$ values represented in a population. This behaviour is due to $\tt{XFit}$'s `supermutation' operator, which introduces brief epochs of higher than usual mutation in a population. The typical width of a population at each generation is influenced by $\tt{XFit}$'s `fuzzy' relative tolerance threshold that identifies individuals within some tolerance of the population's best solution as `optimals', and are given a higher probability of selection. In contrast, the LMA trajectories are spread across a significantly wide range of $\chi_r^2$ values at any given step, indicating that there are many possible unique paths leading to local minima when relying on gradient information to inform updates. Assuming consistent starting values of $\textbf{h}$, $\lambda$, $\Delta_{\text{crit}}$, and $\Delta_{\text{fit}}$ between runs, each LMA fit is deterministic, with trajectories depending only on the initial starting point of the search. Therefore, a small number of LMA trajectories that do end up leading to a good solution implies a relatively small volume of promising initial search points in the parameter space in comparison to those that converge to poor fits, illustrating the LMA's sensitivity to the boundaries of the distribution used to sample the starting point.

The range of optimization steps spanned by a given $\chi_r^2$ also offers clues about the topology of the parameter space. The maximum number of steps taken by the LMA at values of $\chi_r^2 \approx 20$ is approximately 50 steps. Search agents spend significantly less time in this region when compared to $2<\chi_r^2 <15$, where the total number of steps in an optimization can span over 4 orders of magnitude. The gradient-based nature of the LMA suggests that level regions of $\chi_r^2$ in the parameter space containing a sparse number of steep gradients are associated with smaller numbers of optimization steps. This is also demonstrated by the trajectory that results in the best solution found by the LMA, shown as a solid dark-red line in the top panel of Figure \ref{fig:G41-chi-per-gen}. The majority of the paths leading to the best $\chi_r^2$ solutions show a sudden decrease between 40-400 steps before the optimizer quickly converges to its final solution. The LMA's $\chi_r^2$ densities also become less uniformly distributed across each step as the step number in the optimization increases, and comparisons between the top and middle panels of Figure \ref{fig:G41-chi-per-gen} indicate strong correlations between trajectories that result in both small and large $\chi_r^2$ up to a maximum step size (e.g. at steps 40-200,  $6<\chi_r^2<10$ and $2.5<\chi_r^2<3$), at which point the trajectories suddenly diverge. The $\chi_r^2$ densities spanning multiple orders of magnitude between $2.2\leq\chi_r^2 \leq 15$ imply that for certain ranges of $\chi_r^2$ values, there exist hyper-dimensional, level-plateaus in the fitness landscape where the objective function exhibits a decreased sensitivity to changes in the parameters, with one or more relatively small ``funnels'' to lower $\chi_r^2$ regions. LMA performance in these regions is significantly impacted by a correlation between one's choice in convergence criteria and the topology of the fitness landscape, features which generally cannot be known a priori. In a level plateau containing a funnel, smaller values of $\Delta_\text{crit}$ may prove helpful for finding small, sharply declining funnels, by preventing early convergence in the flattened regions surrounding them. However, this invariably comes at the cost of a larger number of fitness function evaluations. On the other hand, the LMA would benefit from a larger $\Delta_{\text{crit}}$ in level plateaus that do not contain a monotonically decreasing path to a smaller $\chi_r^2$, as this would lead to earlier convergence and reserve resources for a greater number of optimization attempts.  

By studying correlations between topological features in the model fitness-landscape and optimizer convergence, we gain a better understanding of the strengths and weaknesses of local and global approaches. Figures \ref{fig:G41-p2-projection}, \ref{fig:G41-p4-projection}, and \ref{fig:G41-p25-projection} depict projections of model parameters across $\chi_r^2$ values and optimizer step (generation), as seen by the LMA (EO) for the parameters $kT_1$, $kT_2$, and $\sigma_7$, respectively. This visualization strategy is identical to those described in Figure \ref{fig:G41-chi-per-gen}. The trajectories carved out by the EO in panel (\textit{iii}) appear to correlate most strongly with the highest density LMA trajectories of panel (\textit{ii}). This is also seen when comparing panels and (\textit{iv}) and (\textit{v}), suggesting that although the EO does not explicitly use gradient information to inform its search, it is nonetheless able to identify and encode these regions when it is beneficial to the optimization process, discarding of this information in favor of encoding other, potentially non-linear, or non-gradient topological features when gradient features are no longer useful.

\begin{figure*}
        \plottwo{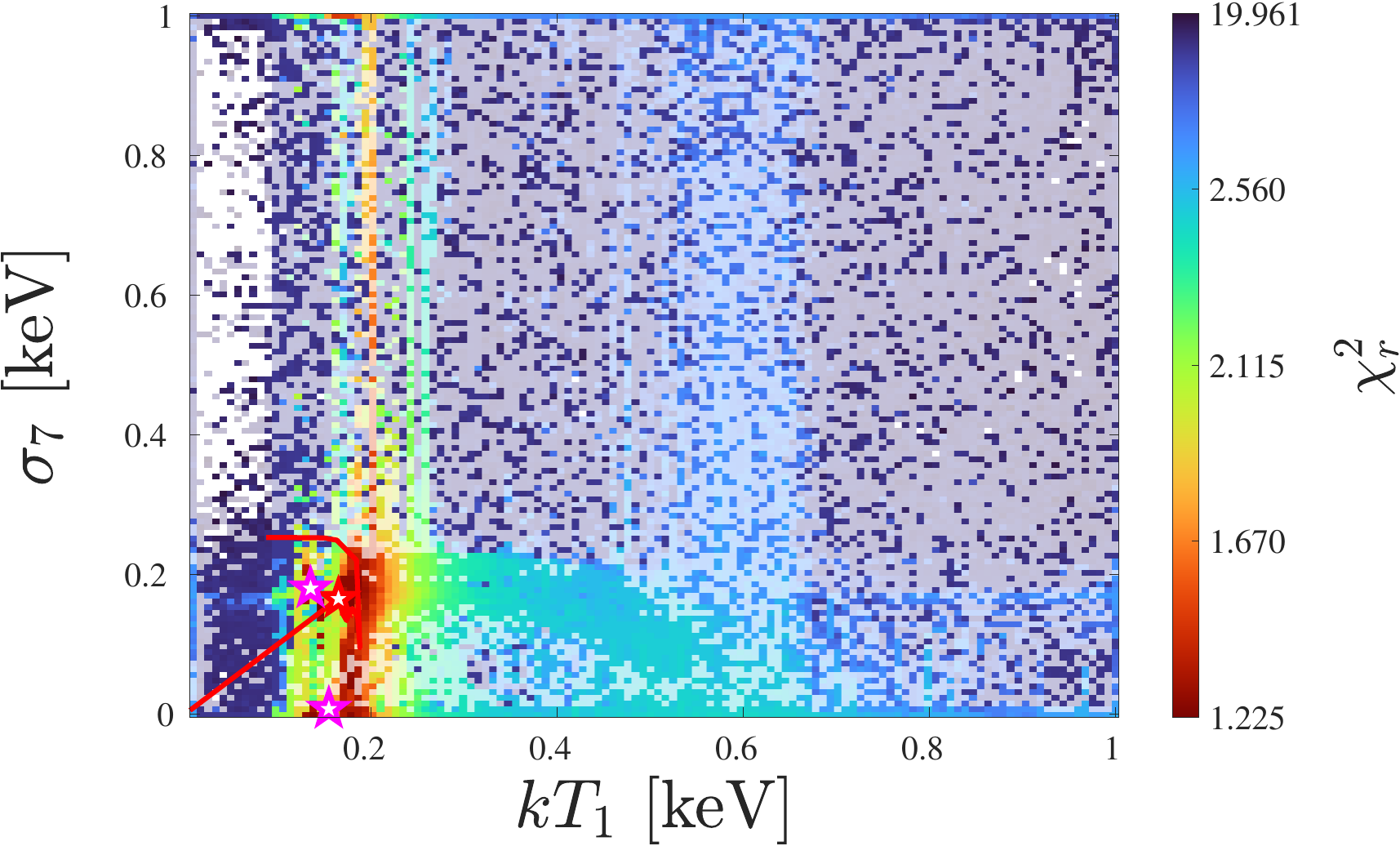}{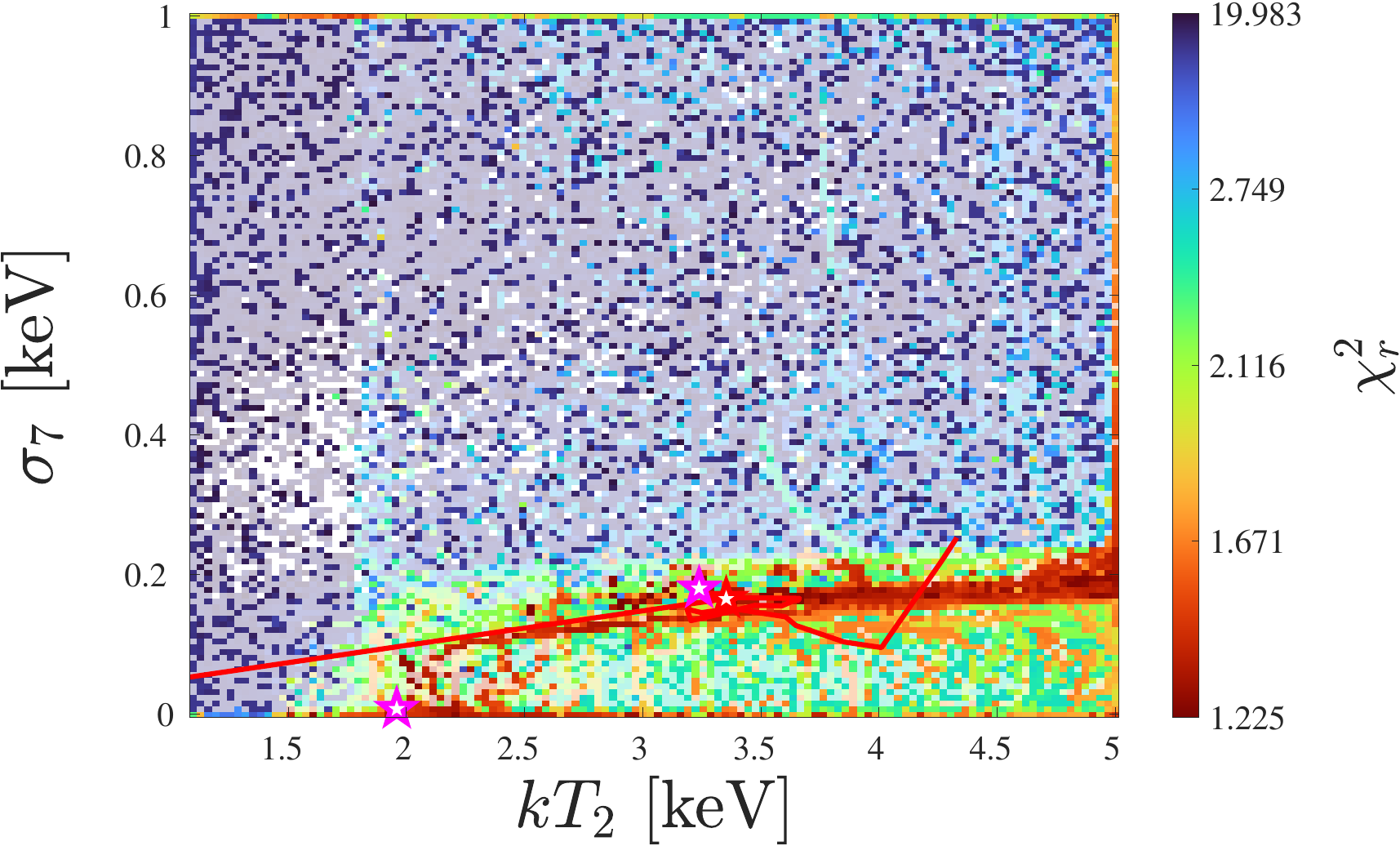}        
    \caption{Parameter space projections of the first and second bremsstrahlung components $kT_1$ and $kT_2$, and the width of the Ca He$\alpha$ line for the absorbed thermal model used to fit the spectrum of G41.1--0.3. The data is spatially binned to include only the best-fit $\chi^2_r$ solution found in each bin and bins of high opacity represent the local minima found across 45297 independent LMA searches. The LMA shows similar trapping behavior when attempting to fit this model as was seen when fitting the five-parameter model of CXOU J232327.9+584842 in Figure \ref{fig:casa-xspec}. In fitting the 29-parameter model, the LMA is corralled down a narrow canyon of solutions shielding Solution 1 ($\chi_r^2=1.238$) from large regions of the parameter space and effectively reducing the number of possible paths capable of reaching it. The solid red line traces the path taken by LMA in finding the Solution 2 ($\chi_r^2=1.225$).\label{fig:g41-xspec-maps-temps-sigma7}}
\end{figure*}

A 2-dimensional projection map of parameter values found by each search attempt made by the LMA for this 29-parameter model is shown in Figure \ref{fig:g41-xspec-maps-temps-sigma7}. Each parameter space map illustrates that the two solutions found for the width of the Ca line component $\sigma_7$ are degenerate with respect to the second temperature component kT$_2$ but not kT$_1$. Such maps also help visualize why certain model components pose significant challenges to the local optimizer. Solution 1 is tucked away in the bottom left corner of the possible range of values for both $kT_2$ and $\sigma_7$ making it statistically much more likely that randomly sampled initial starting point for the search will start out closer to the second solution. Solution 2 is also surrounded by a long narrow valley of smaller $\chi^2$ values that effectively corrals the LMA towards it, shielding the first solution from the local optimizer’s view. Since $\tt{XFit}$ does not rely on gradient information to update its search, it is less susceptible to this trapping behaviour. In the top panel of Figure \ref{fig:G41-p2-projection}, we see that unless the LMA starts its search near $kT_1(\chi_{r}^2=1.225)=0.17$ keV, it tends towards worse $\chi_r^2$ values when it finds itself in the convex basin containing $kT_1(\chi_{r}^2=1.225)$ at large $\chi_r^2$. The explanation for this is illustrated in the left panel of Figure \ref{fig:g41-xspec-maps-temps-sigma7}. This projection shows a large number of local minima separating solution $\chi_r^2=1.225$ from the high-$\chi_r^2$ (dark blue) regions of $0.3<kT_1<1$. Near $kT_1(\chi_{r}^2=1.225)$, there also exist a number of narrow vertical canyons spanning the range $\sigma_7(\chi_{r}^2=1.225)<\sigma_7<1$, showing that the LMA is more likely to find paths to the $3\sigma$ convex basin if it arrives at this $\chi_r^2 \approx1.8$ (orange) canyon. Contrastly, we see no such vertical narrow canyons leading towards the $3\sigma$ convex basin for $kT_2$ in the right panel of Figure \ref{fig:g41-xspec-maps-temps-sigma7}, a feature that is also reflected in the lack of correlation between LMA trajectories for $\chi_r^2<2$ in panels (\textit{i}) and (\textit{ii}) of Figure \ref{fig:G41-p4-projection}. Such features indicate that if the LMA is starting far away from $kT_2$, there is very narrow range of values for $\sigma_7$ that lead to the $3\sigma$ convex basin, further demonstrating the potential pitfalls of performing consecutive local searches within a subspace of the parameters, where although convergence in one parameter may lead to improved fits early on, this may also run the risk of cutting off regions containing better solutions.

Two metrics are defined to allow for comparisons between EO and LMA performance. An \textit{exploration} metric $\mathcal{E}_{\text{ore}}$ emphasizes optimizer effectiveness in mapping regions around solutions in the parameter space over a total number of objective function evaluations $N_{\text{LMA}}$ and $N_{\text{EO}}$ for the local and global optimizers, respectively. Equation \ref{eq:explore} describes  $\mathcal{E}_{\text{ore}}$ as the ratio between the total number of distinct solutions $q(\chi^2_{r,3\sigma})$ found by both optimizers within the convex basin of the 3$\sigma$ confidence interval surrounding solutions 1 and 2. 

\begin{equation}
\mathcal{E}_{\text{ore}}(n)=\frac{q_{\text{LMA}}(\chi_{r,3\sigma}^2)}{N_{\text{LMA}}} \frac{N_{EO}}{q_{\text{EO}}(\chi_{r,3\sigma}^2)}
\label{eq:explore}
\end{equation}

A value of $\mathcal{E_{\text{ore}}}\approx1$ indicates comparable effectiveness in exploratory capability between the optimizers, and $\mathcal{E_{\text{ore}}}<<1$ favors the EO. A solution is only considered distinct and added to $q(\chi_{r,3\sigma}^2)$ if a subsequent step from that solution to any other found by the optimizer would not trigger the stopping criterion set by $\Delta_{\text{crit}}$ and $\Delta_{\text{fit}}$ for the LMA and outlined at the beginning of this section. This condition balances the granularity of the parameter space between the optimizers, excluding solutions found by the EO that are too similar to be distinguished by the LMA and avoids over-counting EO solutions. For a total $N_{EO}$ objective function calls, the EO finds $q_{\text{EO}}(\chi_{r,3\sigma}^2=1.238)=15290$ and $q_{\text{EO}}(\chi_{r,3\sigma}^2=1.225)=843002$ unique solutions in the $3\sigma$ convex basin surrounding Solutions 1 and 2, respectively. In comparison, the LMA finds $q_{\text{LMA}}(\chi_{r,3\sigma}^2=1.238)=39$ and $q_{\text{LMA}}(\chi_{r,3\sigma}^2=1.225)=2713$ solutions across $N_{LMA}=220740499$ objective function evaluations. Values of $\mathcal{E_{\text{ore}}}(\chi_{r,3\sigma}^2=1.238)=2.34\times 10^{-4}$ and $\mathcal{E_{\text{ore}}}(\chi_{r,3\sigma}^2=1.225)=2.96\times10^{-4}$ show strong preference for the EO in terms of its exploratory power.

\begin{figure}
    \plotone{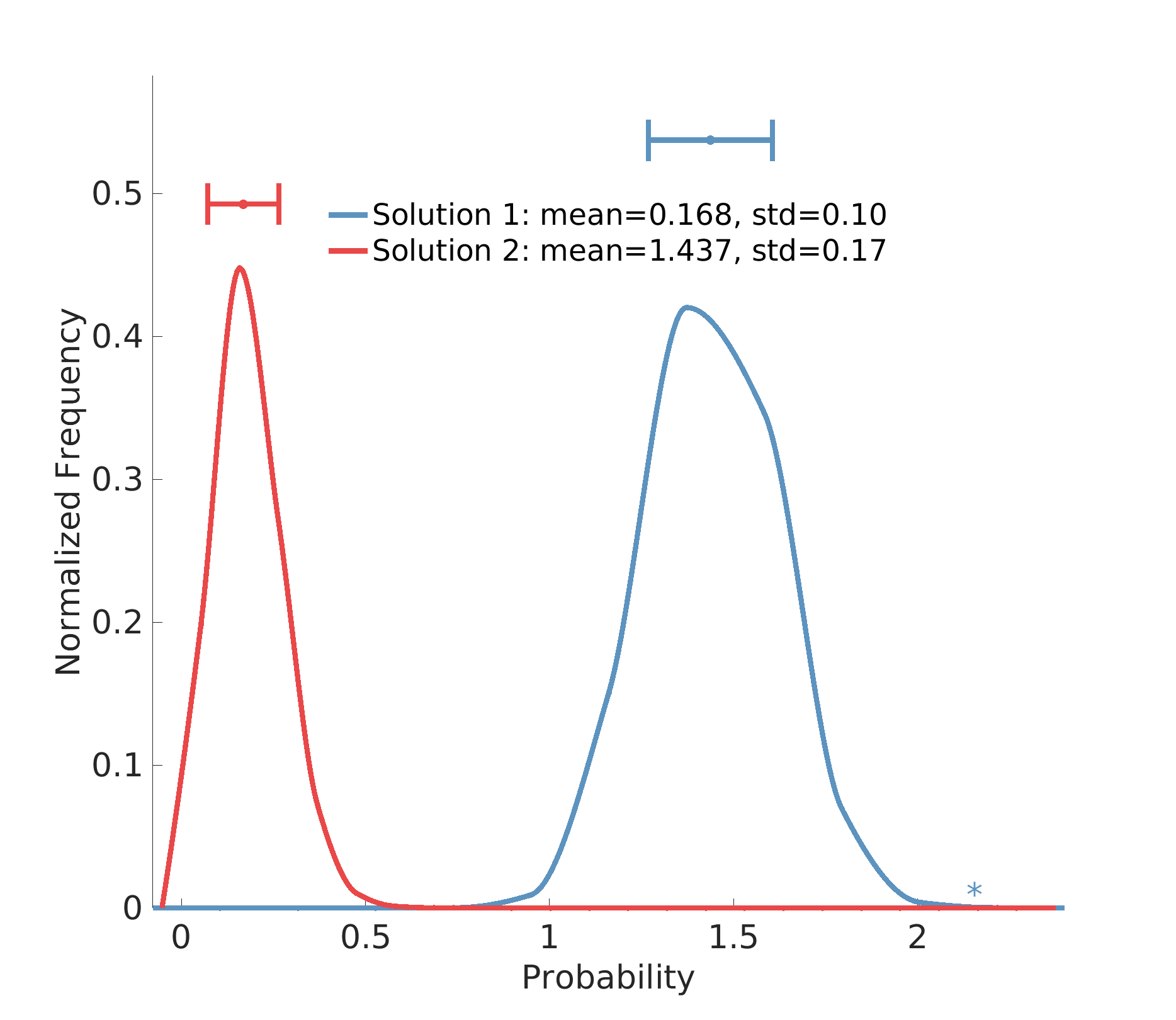}
    \caption{A Monte Carlo simulated probability distribution sampled from $2.1\times 10 ^{5}$ independent LMA runs (totaling $\approx 2.2 \times 10^{8}$ objective function evaluations). The distribution was generated using $10^5$ simulations and finds a $\mathcal{E}_{\text{oit}}(\chi_r^2=1.238)=0.17$ and $\mathcal{E}_{\text{oit}}(\chi_r^2=1.225)=1.44$ relative probability for the LMA to converge to degenerate Solutions 1 and 2 respectively given $\text{N}_{\text{G41}}=20285430$ objective function evaluations (equal to that of the EO).\label{fig:MC}}   
\end{figure}

An \textit{exploitation} metric $\mathcal{E}_{\text{oit}}$ emphasizes optimizer efficiency and measures the smallest number of fitness function evaluations required by each optimizer to converge to $\chi_{r,3\sigma}^2$ for each solution. The EO converges to Solutions 1 and 2 in $N_{EO}(\chi_{r,3\sigma}^2=1.238)=12355392$ and $N_{EO}(\chi_{r,3\sigma}^2=1.225)=4349990$ objective function evaluations, respectively. $10^5$ Monte Carlo simulations were performed for the LMA across a total of $N_{LMA}(\chi_{r,3\sigma}^2)=220739152$ objective function evaluations for each simulation. To generate the probability distribution shown in Figure \ref{fig:MC}, statistics from each LMA attempt were collected only up to and including the step that converged to the $\chi_{r,3\sigma}^2$ basin. The number of LMA attempts that successfully converged within the $\chi_{r,3\sigma}^2$ basin are given by $\text{m}(\chi_{r,3\sigma}^2=1.238)=3$ and $\text{m}(\chi_{r,3\sigma}^2=1.225)=73$ for solutions 1 and 2, respectively. The relative probability of the LMA converging to $\chi_{r,3\sigma}^2$ within the allotted $N_{EO}(\chi_{r,3\sigma}^2)$ for each solution is given by Equation \ref{eq:exploit}.
\begin{equation}
\mathcal{E}_{\text{oit}}(\chi_{r,3\sigma}^2)=\frac{\text{m}_{LMA}(\chi_{r,3\sigma}^2)}{N_{LMA}(\chi_{r,3\sigma}^2)} N_{EO}(\chi_{r,3\sigma}^2) 
\label{eq:exploit}
\end{equation}

\noindent The probability of the LMA finding Solution 1 is $\mathcal{E}_{\text{oit}}(\chi_{r,3\sigma}^2=1.238)= 0.17$ when compared to the EO for an equal number of objective function evaluations, showing strong preference for the EO in terms of exploitative efficiency. In contrast, $\mathcal{E}_{\text{oit}}(\chi_{r,3\sigma}^2=1.225)=1.44$ indicates that the local and global optimizers are comparable in this metric for solution 2.  

\begin{figure*}
        \plotone{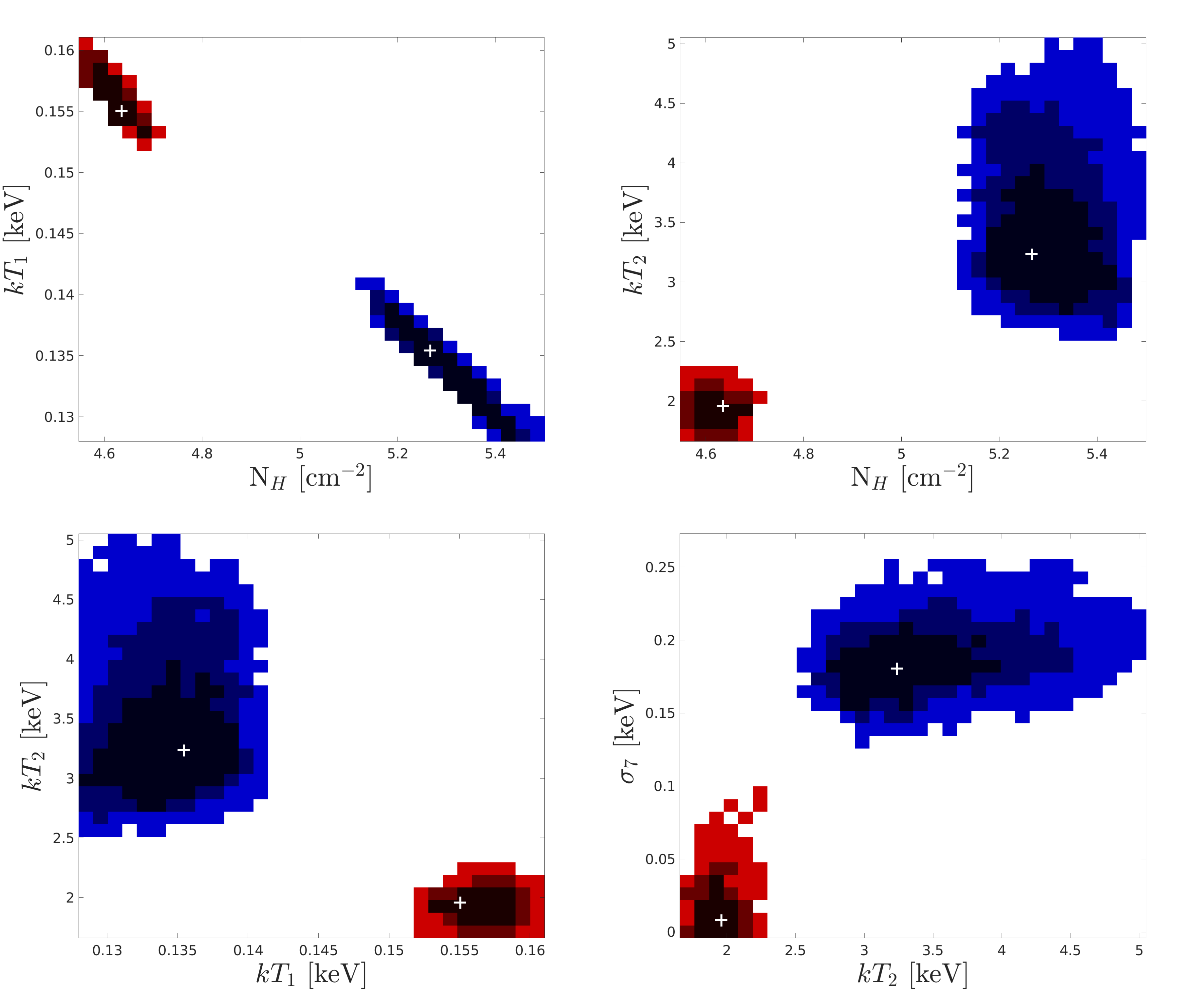}
        \caption{The 1$\sigma$, 2$\sigma$, and 3$\sigma$ confidence regions automatically mapped by $\tt{XFit}$ during optimization. The parameter uncertainties of Solution 1 (red) occupy a smaller volume in the temperature components $kT_1$ and $kT_2$ relative to Solution 2 (blue), suggesting that solutions associated with a greater degree of uncertainty in their parameters represent larger targets for the optimizer with a higher probability of being found.\label{fig:G41-xfit-maps-degen}}
\end{figure*}

Figure \ref{fig:G41-xfit-maps-degen} shows two distinct solutions as islands of 3$\sigma$ confidence whose values are used to calculate the uncertainties given in Table \ref{table:g41-table}. Had this problem been solved by conventional means, one of these solutions may have been missed entirely. Compared with Solution 2, Solution 1 is clearly more challenging for both optimizers to find based on the number of fitness function evaluations required to converge to the $\chi_{r,3\sigma}^2=1.238$ convex basin, and the number of distinct solutions mapped for each region. This is partially explained by the difference in the parameter space volume filled by the $3\sigma$ convex basin surrounding each solution. Solution 2 (blue) is associated with a higher degree of uncertainty in its parameters, representing a much larger target for an optimizer. In the case of the LMA, a larger target is more likely to contain a greater number of trajectories, and consequently, a greater number of initial starting points, that lead to it. The $3\sigma$ confidence volume of Solution 2 influences the performance of the EO as well, since there are a greater number of representations available to it within the objective function's fuzzy tolerance. This suggests that the greater the precision in the parameter values predicted by a model's solution, the harder the solution is to find, particularly when it resides in a parameter space containing other, degenerate solutions with a greater degree of uncertainty. 

Despite being unable to find the second degenerate solution within the allotted number of objective function evaluations, the LMA demonstrates a particular utility in mapping potentially problematic regions of the parameter space when compared with the EO, emphasizing a promising synergy when utilizing local and global optimization methods together in identifying useful metrics for optimizer performance, with the goal of improving the robustness of local and global optimization algorithms alike.

\section{Conclusions}
We have developed and extensively tested $\tt{XFit}$, an X-ray spectral fitting code written in MATLAB, that makes use of the Ferret EO, which is part of the \textit{Qubist} Optimization Toolbox. The \textit{Qubist} package provides a variety of local optimization tools in addition to the global \textit{Ferret} EO. These tools allow for automated search and exploration of spectral model parameter spaces, requiring minimal input from the user to locate best-fit solutions.

In this work we analysed a simple, five-parameter absorbed thermal model applied to the CCO in Cassiopeia A (CXOU J232327.9+584842), and a more complicated, 29-parameter Gaussian line model applied to the SNR G41.1--0.3, as representative examples to demonstrate cases of spectral fitting where global optimization methods are needed. Our fully automated and parallel search method illuminates the parameter space structure of the individual fits, providing insight into the efficacy of optimization techniques used to explore solutions in fitness landscapes. This is made possible by the mapping features and advanced exploration capabilities of the Ferret Evolutionary Optimizer, which return families of optimal solutions that occupy the $1\sigma$, $2\sigma$ and $3\sigma$ confidence intervals during a run (Figures \ref{fig:casa-conf-grid} and \ref{fig:G41-xfit-maps-degen}). This is in contrast with the local, single-search-agent, gradient-based LMA, that is designed to return a single best-fit solution. $\tt{XFit}$s optimal solutions are robust and repeatable between individual runs of the $\tt{XFit}$ code. This global approach is extremely valuable for analyzing the behavior of model parameters when fitting data, and is especially useful for the thorough exploration and mapping of high-dimensional model parameter spaces (Figures \ref{fig:casa-xspec} and \ref{fig:g41-xspec-maps-temps-sigma7}). Multiple performance metrics: $\mathcal{E}_{\text{ore}}$ and $\mathcal{E}_{\text{oit}}$ are defined, which compare the exploratory effectiveness and exploitative efficiency of both optimizers, respectively. Comparisons between the local LMA and global EO methods are found to favour global methods as models increase in number of parameters and search-space hypervolume, due to their strong discrimination of large-$\chi_r^2$ regions, and their ability to encode complex features in the parameter space, which are shared usefully among individuals in a population. Novel visualizations are introduced in Figure \ref{fig:G41-chi-per-gen}, and Figures \ref{fig:G41-p2-projection}, \ref{fig:G41-p4-projection}, and \ref{fig:G41-p25-projection} in the appendix, as a means of making comparisons between the convergence behaviour of different optimization methods.

Although a single LMA search tends to require a smaller number of objective function evaluations than the EO to converge to a local minimum, it is unable to consistently find and map solutions when these approaches are compared using the above performance metrics. Our results suggest that in addition to considering the number of objective function evaluations required for updating the search coordinates of an optimization, other metrics should also be considered when deciding between local and global approaches for model fitting. We also show that LMA performance is more sensitive to the initial starting point of the search when compared with the EO, as $\tt{XFit}$ demonstrates more robustness against trapping by local minima. Local optimization techniques typically require researchers to make progress on a fitting problem by holding certain parameters constant and sequentially performing fits over subspaces of parameter values. Based on our results, it is not clear that such an approach is sufficiently robust for large model parameter spaces, and can increase the likelihood of introducing bias into the resulting best-fit solutions that are used to infer properties associated with astrophysical processes. In such cases, $\tt{XFit}$ reduces the need for human intervention in optimization and decreases the risk of introducing bias into the resulting fit parameters.

In summary, we present $\tt{XFit}$, a new spectral fitting tool that is particularly robust for high-dimensional problems and capable of uncovering solutions that the Levenberg–Marquardt algorithm may miss. We demonstrate and emphasize that $\tt{XFit}$ is designed as a complimentary approach to the $\tt{XSPEC}$ software, rather than a replacement, and makes use of $\tt{XSPEC}$'s library of spectral models. This development is especially timely given the surge of high-resolution X-ray spectroscopy data from XRISM and the upcoming NewAthena mission, and future work will explore $\tt{XFit}$’s performance on larger parameter spaces and broader datasets. The authors plan to release a publicly available version of $\tt{XFit}$ in the near future. Researchers interested in using $\tt{XFit}$ should contact the primary author.
\\

This research is primarily supported by the Natural Sciences and Engineering Research Council of Canada (NSERC) through the Discovery Grants and Canada Research Chairs Programs (S.S.H.).
This research made use of NASA's Astrophysics Data System and the HEASARC database.

\software{Qubist \citep{fiege-2010}, MATLAB \citep{MATLAB}, $\tt{XSPEC}$ \citep{Arnaud1996}, Plot and compare histograms; pretty by default \citep{MCPlot}, 200 colormap \citep{200cm}} 

\appendix

\begin{figure*}
        \epsscale{0.97}
        \plotone{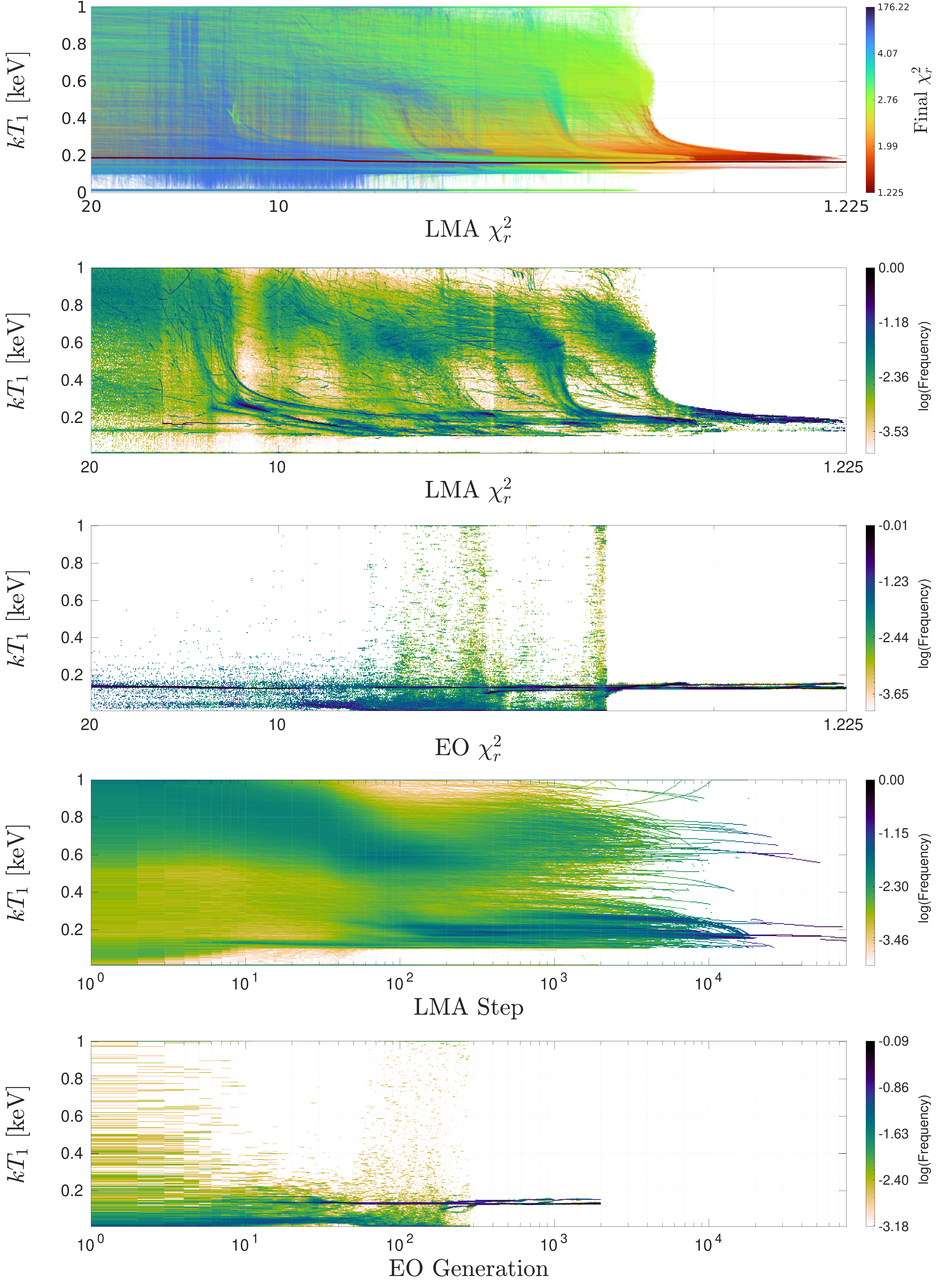}
        \caption{Parameter space projections for the temperature of the first bremsstrahlung component kT$_1$. In descending  order from the topmost figure: \textit{(i)} LMA trajectories through the parameter space for each independent optimization attempt as a function of the monotonically decreasing objective function value using the same color-coding as the top Panel of Figure \ref{fig:G41-chi-per-gen}. \textit{(ii)} LMA $\chi_r^2$ density map as a function of parameter value and color-coded identically to that of the middle panel of Figure \ref{fig:G41-chi-per-gen}. \textit{(iii)} EO $\chi_r^2$ density map as a function of parameter value and color-coded identically to that of the bottom panel of Figure \ref{fig:G41-chi-per-gen}.  \textit{(iv)} Same as \textit{(ii)} but projected as a function of optimizer step. \textit{(v)} Same as \textit{(iii)} but projected as a function of optimizer generation.\label{fig:G41-p2-projection}}
\end{figure*}

\begin{figure*}
        \epsscale{0.97}
        \plotone{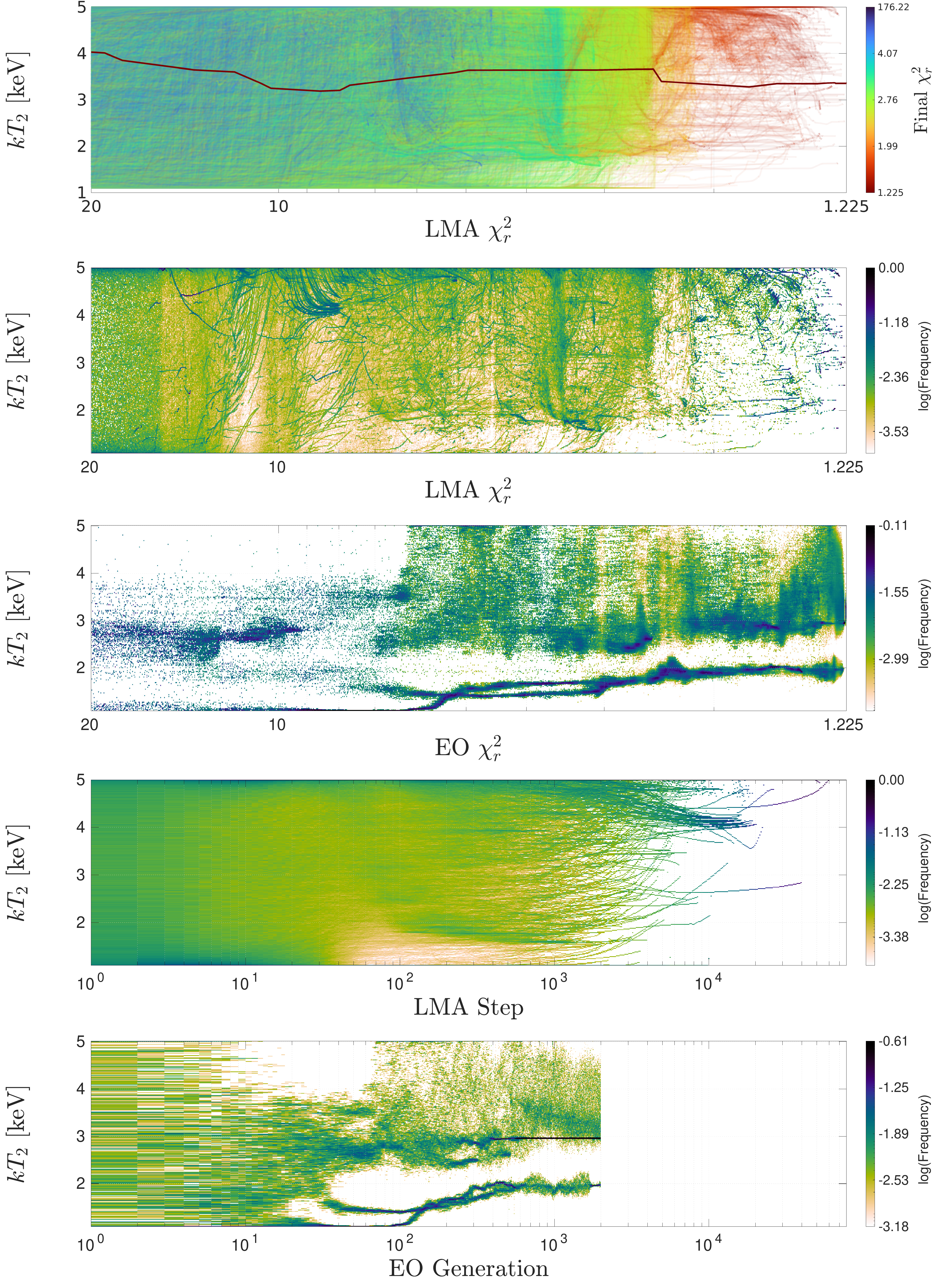}      
        \caption{Parameter space projections for the temperature of the second bremsstrahlung component kT$_2$. See Figure \ref{fig:G41-p2-projection} for descriptions of each panel.\label{fig:G41-p4-projection}}
\end{figure*}

\begin{figure*}
        \epsscale{0.97}
        \plotone{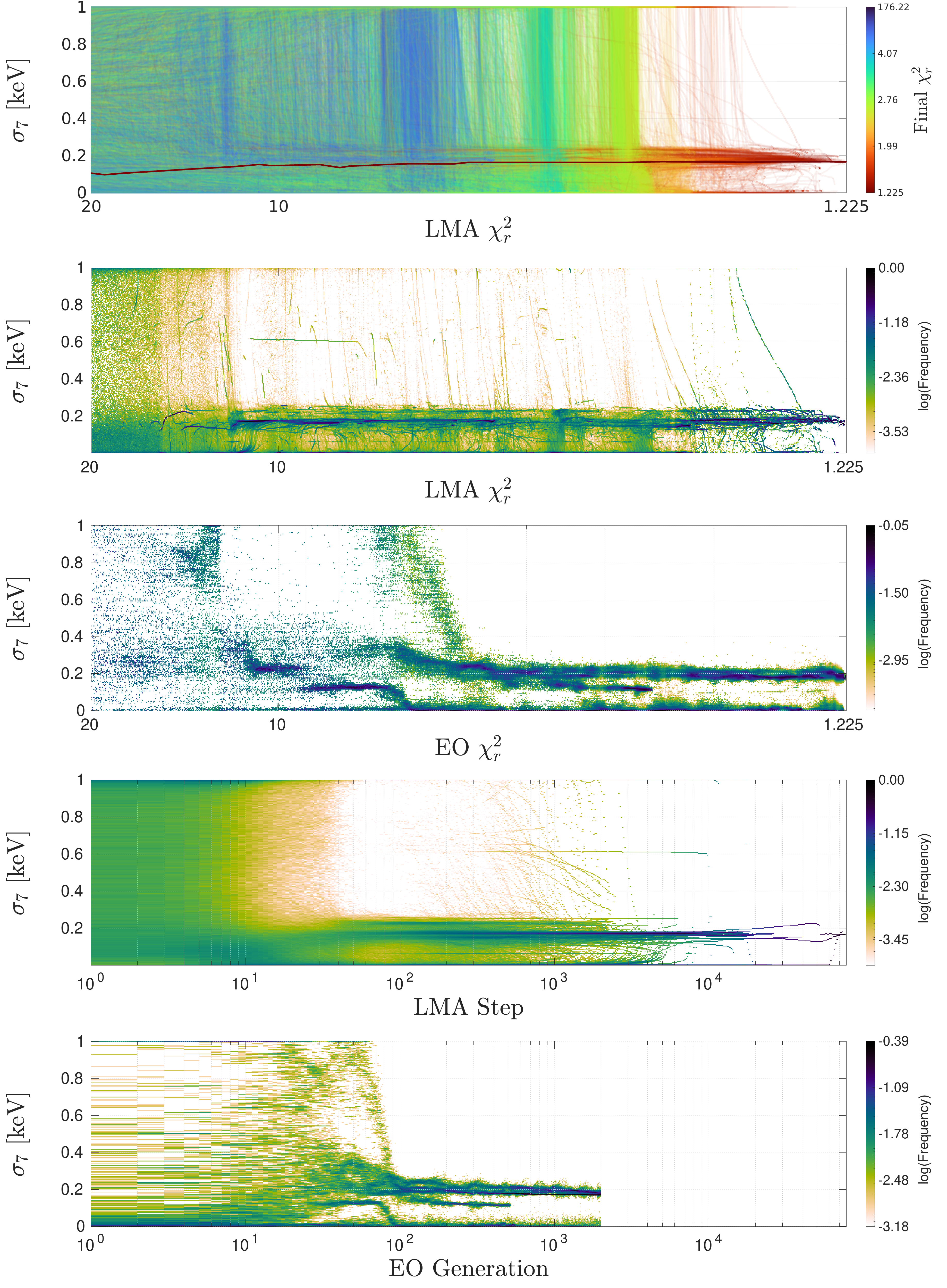}       
        \caption{Parameter space projections for the line width of the seventh Gaussian line component $\sigma_7$. See Figure \ref{fig:G41-p2-projection} for descriptions of each panel.\label{fig:G41-p25-projection}}
\end{figure*}

\bibliography{main}{}

@INPROCEEDINGS{2008AIPC..983..171K,
       author = {{Kargaltsev}, O. and {Pavlov}, G.~G.},
        title = "{Pulsar Wind Nebulae in the Chandra Era}",
    booktitle = {40 Years of Pulsars: Millisecond Pulsars, Magnetars and More},
         year = 2008,
       editor = {{Bassa}, C. and {Wang}, Z. and {Cumming}, A. and {Kaspi}, V.~M.},
       series = {American Institute of Physics Conference Series},
       volume = {983},
        month = feb,
    publisher = {AIP},
        pages = {171-185},
          doi = {10.1063/1.2900138},
archivePrefix = {arXiv},
       eprint = {0801.2602},
 primaryClass = {astro-ph},
       adsurl = {https://ui.adsabs.harvard.edu/abs/2008AIPC..983..171K}
}

@ARTICLE{2008ARA&A..46...89R,
       author = {{Reynolds}, S.~P.},
        title = "{Supernova remnants at high energy.}",
      journal = {\araa},
         year = 2008,
        month = sep,
       volume = {46},
        pages = {89-126},
          doi = {10.1146/annurev.astro.46.060407.145237},
       adsurl = {https://ui.adsabs.harvard.edu/abs/2008ARA&A..46...89R}
}

@ARTICLE{2018ApJ...862L..19N,
       author = {{Nynka}, Melania and {Ruan}, John J. and {Haggard}, Daryl and {Evans}, Phil A.},
        title = "{Fading of the X-Ray Afterglow of Neutron Star Merger GW170817/GRB 170817A at 260 Days}",
      journal = {\apjl},
         year = 2018,
        month = aug,
       volume = {862},
       number = {2},
          eid = {L19},
        pages = {L19},
          doi = {10.3847/2041-8213/aad32d},
archivePrefix = {arXiv},
       eprint = {1805.04093},
 primaryClass = {astro-ph.HE},
       adsurl = {https://ui.adsabs.harvard.edu/abs/2018ApJ...862L..19N}
}

@ARTICLE{2000ApJ...531L..53P,
       author = {{Pavlov}, G.~G. and {Zavlin}, V.~E. and {Aschenbach}, B. and {Tr{\"u}mper}, J. and {Sanwal}, D.},
        title = "{The Compact Central Object in Cassiopeia A: A Neutron Star with Hot Polar Caps or a Black Hole?}",
      journal = {\apjl},
         year = 2000,
        month = mar,
       volume = {531},
       number = {1},
        pages = {L53-L56},
          doi = {10.1086/312521},
archivePrefix = {arXiv},
       eprint = {astro-ph/9912024},
 primaryClass = {astro-ph},
       adsurl = {https://ui.adsabs.harvard.edu/abs/2000ApJ...531L..53P}
}

@ARTICLE{2013ApJ...765...58G,
       author = {{Gotthelf}, E.~V. and {Halpern}, J.~P. and {Alford}, J.},
        title = "{The Spin-down of PSR J0821-4300 and PSR J1210-5226: Confirmation of Central Compact Objects as Anti-magnetars}",
      journal = {\apj},
         year = 2013,
        month = mar,
       volume = {765},
       number = {1},
          eid = {58},
        pages = {58},
          doi = {10.1088/0004-637X/765/1/58},
archivePrefix = {arXiv},
       eprint = {1301.2717},
 primaryClass = {astro-ph.HE},
       adsurl = {https://ui.adsabs.harvard.edu/abs/2013ApJ...765...58G}
}

@ARTICLE{2023MNRAS.525.6257B,
       author = {{Braun}, C. and {Safi-Harb}, S. and {Fryer}, C.~L. and {Zhou}, P.},
        title = "{Progenitors and explosion properties of supernova remnants hosting central compact objects: II. A global systematic study with a comparison to nucleosynthesis models}",
      journal = {\mnras},
         year = 2023,
        month = nov,
       volume = {525},
       number = {4},
        pages = {6257-6284},
          doi = {10.1093/mnras/stad2592},
archivePrefix = {arXiv},
       eprint = {2308.13693},
 primaryClass = {astro-ph.HE},
       adsurl = {https://ui.adsabs.harvard.edu/abs/2023MNRAS.525.6257B}
}

@ARTICLE{Giacconi1962,
       author = {{Giacconi}, Riccardo and others},
        title = "{Evidence for x Rays From Sources Outside the Solar System}",
      journal = {\prl},
         year = 1962,
        month = dec,
       volume = {9},
       number = {11},
        pages = {439-443},
          doi = {10.1103/PhysRevLett.9.439},
       adsurl = {https://ui.adsabs.harvard.edu/abs/1962PhRvL...9..439G}
}

@INPROCEEDINGS{Brickhouse2000,
       author = {{Brickhouse}, N.~S.},
        title = "{Chandra and XMM-Newton: Atomic Data Needs for X-ray Astronomy}",
    booktitle = {IAU Joint Discussion},
         year = 2000,
       series = {IAU Joint Discussion},
       volume = {24},
        month = jan,
          eid = {19},
        pages = {19},
       adsurl = {https://ui.adsabs.harvard.edu/abs/2000IAUJD...1E..19B}
}

@ARTICLE{Hitomi2018,
       author = {{Hitomi Collaboration} and others},
        title = "{Atomic data and spectral modeling constraints from high-resolution X-ray observations of the Perseus cluster with Hitomi}",
      journal = {\pasj},
         year = 2018,
        month = mar,
       volume = {70},
       number = {2},
          eid = {12},
        pages = {12},
          doi = {10.1093/pasj/psx156},
archivePrefix = {arXiv},
       eprint = {1712.05407},
 primaryClass = {astro-ph.HE},
       adsurl = {https://ui.adsabs.harvard.edu/abs/2018PASJ...70...12H}}

@INPROCEEDINGS{Arnaud1996,
       author = {{Arnaud}, K.~A.},
        title = "{XSPEC: The First Ten Years}",
    booktitle = {Astronomical Data Analysis Software and Systems V},
         year = 1996,
       editor = {{Jacoby}, George H. and {Barnes}, Jeannette},
       series = {Astronomical Society of the Pacific Conference Series},
       volume = {101},
        month = jan,
        pages = {17},
       adsurl = {https://ui.adsabs.harvard.edu/abs/1996ASPC..101...17A}
}

@BOOK{Bevington2003,
       author = {{Bevington}, Philip R. and {Robinson}, D. Keith},
        title = "{Data reduction and error analysis for the physical sciences}",
         year = 2003,
        publisher = {McGraw-Hill},
       adsurl = {https://ui.adsabs.harvard.edu/abs/2003drea.book.....B}
}

@BOOK{Holland1975,
       author = {{Holland}, John H.},
         title = "{Adaptation in natural and artificial systems. an introductory analysis with applications to biology, control and artificial intelligence}",
         year = 1975,
    publisher={The MIT Press},
       adsurl = {https://ui.adsabs.harvard.edu/abs/1975anas.book.....H}
}

@BOOK{Goldberg1989,
       author = {{Goldberg}, David E.},
        title = "{Genetic algorithms in search, optimization and machine learning}",
    publisher={Addison-Wesley Publishing Company, Inc.},
         year = 1989,
       adsurl = {https://ui.adsabs.harvard.edu/abs/1989gaso.book.....G}
}

@ARTICLE{Charbonneau1995,
       author = {{Charbonneau}, P.},
        title = "{Genetic Algorithms in Astronomy and Astrophysics}",
      journal = {\apjs},
         year = 1995,
        month = dec,
       volume = {101},
        pages = {309},
          doi = {10.1086/192242},
       adsurl = {https://ui.adsabs.harvard.edu/abs/1995ApJS..101..309C}
}

@ARTICLE{fiege-2010,
       author = {{Fiege}, J. D.},
        title = "{Qubist Users Guide: Optimization, Data
Modeling, and Visualization with the Qubist Optimization Toolbox for MATLAB}",
      journal = {(Winnipeg Canada: nQube
Technical Computing)},     
         year = 2025,
}

@ARTICLE{Tananbaum1999,
       author = {{Tananbaum}, H.},
        title = "{Cassiopeia A}",
      journal = {\iaucirc},
         year = 1999,
        month = sep,
       volume = {7246},
        pages = {1},
       adsurl = {https://ui.adsabs.harvard.edu/abs/1999IAUC.7246....1T}
}

@article{heyl_2001, title={The Central X-Ray Point Source in Cassiopeia A}, volume={546}, number={2}, DOI={10.1086/318994}, journal={The Astrophyscial Journal}, author={Heyl, J. S. and others}, year={2001}, month={Feb}, pages={800-810}}

@article{deluca_2017, title={Central Compact Objects in Supernova Remnants}, volume={932}, number={1}, DOI={10.1088/1742-6596/932/1/012006}, journal={Journal of Physics: Conference Series}, author={De Luca, A}, year={2017}, month={Dec}, pages={012006}}

@article{pavlov_2001, title={The Compact Central Object in the RX J0852-4622 Supernova Remnant}, volume={559}, number={2}, DOI={10.1086/323975}, journal={The Astrophyscial Journal}, author={Pavlov, G. G. and Sanwal, D. and Kiziltan, B. and Garmire, G. P.}, year={2001}, month={Oct}, pages={L131-L134}}

@article{pavlov_2004, title={Central Central Objects in Supernova Remnants}, volume={218}, DOI={10.48550/arXiv.astro-ph/0311526}, journal={Young Neutron Stars and Their Environments, IAU Symposium}, author={Pavlov, G. G. and Sanwal, D. and Teter, M. A.}, year={2004}, month={July}, pages={239-246}}

@article{ozel_2016, title={Masses, Radii, and the Equation of State of Neutron Stars}, volume={54}, DOI={10.1146/annurev-astro-081915-023322}, journal={Annual Review of Astronomy and Astrophysics}, author={Ozel, P. and Freire, P.}, year={2016}, month={September}, pages={401-440}}

@article{pavlov_2009, title={A Dedicated Chandra ACIS Observation of the Central Compact Object in the Cassieopia A Supernova Remnant}, volume={703}, number={1}, DOI={10.1088/0004-637X/703/1/910}, journal={The Astrophysical Journal}, author={Pavlov, G. G. and Luna, G. J. M.}, year={2009}, month={Sept}, pages={910-921}}

@ARTICLE{Safi-Harb2000,
       author = {{Safi-Harb}, S. and others},
        title = "{A Broadband X-Ray Study of Supernova Remnant 3C 397}",
      journal = {\apj},
         year = 2000,
        month = dec,
       volume = {545},
       number = {2},
        pages = {922-938},
          doi = {10.1086/317823},
archivePrefix = {arXiv},
       eprint = {astro-ph/0007372},
 primaryClass = {astro-ph},
       adsurl = {https://ui.adsabs.harvard.edu/abs/2000ApJ...545..922S}
}

@ARTICLE{Safi-Harb2005,
       author = {{Safi-Harb}, S. and others},
        title = "{Chandra Spatially Resolved Spectroscopic Study and Multiwavelength Imaging of the Supernova Remnant 3C 397 (G41.1-0.3)}",
      journal = {\apj},
         year = 2005,
        month = jan,
       volume = {618},
       number = {1},
        pages = {321-338},
          doi = {10.1086/425960},
archivePrefix = {arXiv},
       eprint = {astro-ph/0407121},
 primaryClass = {astro-ph},
       adsurl = {https://ui.adsabs.harvard.edu/abs/2005ApJ...618..321S}
}

@ARTICLE{Green2004,
       author = {{Green}, D. A.},
        title = "{A Catalogue of Galactic Supernova Remnants (2004 January version)}",
      journal = {Mullard Radio Astronomy Observatory, Cavendish Laboratory, Cambridge, United Kingdom (available at "http://www.mrao.cam.ac.uk/surveys/snrs/")},
       year = 2004    
}

@misc{MATLAB,
year = {2023},
author = {{The MathWorks Inc.}},
title = {MATLAB version: 9.13.0 (R2023b)},
publisher = {The MathWorks Inc.},
address = {Natick, Massachusetts, United States},
url = {https://www.mathworks.com}
}

@ARTICLE{Wilms2000,
       author = {{Wilms}, J. and {Allen}, A. and {McCray}, R.},
        title = "{On the Absorption of X-Rays in the Interstellar Medium}",
      journal = {\apj},
         year = 2000,
        month = oct,
       volume = {542},
       number = {2},
        pages = {914-924},
          doi = {10.1086/317016},
archivePrefix = {arXiv},
       eprint = {astro-ph/0008425},
 primaryClass = {astro-ph},
       adsurl = {https://ui.adsabs.harvard.edu/abs/2000ApJ...542..914W}
}

@ARTICLE{Balucinska-Church1992,
       author = {{Balucinska-Church}, Monika and {McCammon}, Dan},
        title = "{Photoelectric Absorption Cross Sections with Variable Abundances}",
      journal = {\apj},
         year = 1992,
        month = dec,
       volume = {400},
        pages = {699},
          doi = {10.1086/172032},
       adsurl = {https://ui.adsabs.harvard.edu/abs/1992ApJ...400..699B}
}

@ARTICLE{Morrison1983,
       author = {{Morrison}, R. and {McCammon}, D.},
        title = "{Interstellar photoelectric absorption cross sections, 0.03-10 keV.}",
      journal = {\apj},
         year = 1983,
        month = jul,
       volume = {270},
        pages = {119-122},
          doi = {10.1086/161102},
       adsurl = {https://ui.adsabs.harvard.edu/abs/1983ApJ...270..119M}
}

@ARTICLE{Anders1989,
       author = {{Anders}, E. and {Grevesse}, N.},
        title = "{Abundances of the elements: Meteoritic and solar}",
      journal = {\gca},
         year = 1989,
        month = jan,
       volume = {53},
       number = {1},
        pages = {197-214},
          doi = {10.1016/0016-7037(89)90286-X},
       adsurl = {https://ui.adsabs.harvard.edu/abs/1989GeCoA..53..197A}
}

@article{Levenberg1944,
 ISSN = {0033569X, 15524485},
 URL = {http://www.jstor.org/stable/43633451},
 author = {KENNETH LEVENBERG},
 journal = {Quarterly of Applied Mathematics},
 number = {2},
 pages = {164--168},
 publisher = {Brown University},
 title = {A METHOD FOR THE SOLUTION OF CERTAIN NON-LINEAR PROBLEMS IN LEAST SQUARES},
 urldate = {2024-08-29},
 volume = {2},
 year = {1944}
}

@ARTICLE{Weinberg2013,
       author = {{Weinberg}, David H. and {Mortonson}, Michael J. and {Eisenstein}, Daniel J. and {Hirata}, Christopher and {Riess}, Adam G. and {Rozo}, Eduardo},
        title = "{Observational probes of cosmic acceleration}",
      journal = {\physrep},
         year = 2013,
        month = sep,
       volume = {530},
       number = {2},
        pages = {87-255},
          doi = {10.1016/j.physrep.2013.05.001},
archivePrefix = {arXiv},
       eprint = {1201.2434},
 primaryClass = {astro-ph.CO},
       adsurl = {https://ui.adsabs.harvard.edu/abs/2013PhR...530...87W}
}

@ARTICLE{Heuer2021,
       author = {{Heuer}, Keri and {Foster}, Adam R. and {Smith}, Randall},
        title = "{Spectral Implications of Atomic Uncertainties in Optically Thin Hot Plasmas}",
      journal = {\apj},
         year = 2021,
        month = feb,
       volume = {908},
       number = {1},
          eid = {3},
        pages = {3},
          doi = {10.3847/1538-4357/abcaff},
archivePrefix = {arXiv},
       eprint = {2011.08230},
 primaryClass = {astro-ph.HE},
       adsurl = {https://ui.adsabs.harvard.edu/abs/2021ApJ...908....3H}
}

@ARTICLE{Mukai2003,
       author = {{Mukai}, K. and {Kinkhabwala}, A. and {Peterson}, J.~R. and {Kahn}, S.~M. and {Paerels}, F.},
        title = "{Two Types of X-Ray Spectra in Cataclysmic Variables}",
      journal = {\apjl},
         year = 2003,
        month = mar,
       volume = {586},
       number = {1},
        pages = {L77-L80},
          doi = {10.1086/374583},
archivePrefix = {arXiv},
       eprint = {astro-ph/0301557},
 primaryClass = {astro-ph},
       adsurl = {https://ui.adsabs.harvard.edu/abs/2003ApJ...586L..77M}
}

@ARTICLE{Raga2002,
       author = {{Raga}, A.~C. and {Noriega-Crespo}, A. and {Vel{\'a}zquez}, P.~F.},
        title = "{The X-Ray Luminosities of Herbig-Haro Objects}",
      journal = {\apjl},
         year = 2002,
        month = sep,
       volume = {576},
       number = {2},
        pages = {L149-L152},
          doi = {10.1086/343760},
archivePrefix = {arXiv},
       eprint = {astro-ph/0208079},
 primaryClass = {astro-ph},
       adsurl = {https://ui.adsabs.harvard.edu/abs/2002ApJ...576L.149R}
}

@ARTICLE{Silk1978,
       author = {{Silk}, J. and {White}, S.~D.~M.},
        title = "{The determination of q$_{0}$ using X-ray and microwave observation of galaxy clusters.}",
      journal = {\apjl},
         year = 1978,
        month = dec,
       volume = {226},
        pages = {L103-L106},
          doi = {10.1086/182841},
       adsurl = {https://ui.adsabs.harvard.edu/abs/1978ApJ...226L.103S}
}

@ARTICLE{Safi-Harb2019,
       author = {{Safi-Harb}, Samar and {Doerksen}, Neil and {Rogers}, Adam and {Fryer}, Chris L.},
        title = "{Chandra X-Ray Observations of the Neutron Star Merger GW170817: Thermal X-Ray Emission From a Kilonova Remnant?}",
      journal = {\jrasc},
         year = 2019,
        month = feb,
       volume = {113},
       number = {1},
        pages = {7},
          doi = {10.48550/arXiv.1812.11320},
archivePrefix = {arXiv},
       eprint = {1812.11320},
 primaryClass = {astro-ph.HE},
       adsurl = {https://ui.adsabs.harvard.edu/abs/2019JRASC.113....7S}
}

@ARTICLE{Yamaguchi2014,
       author = {{Yamaguchi}, Hiroya and {Badenes}, Carles and {Petre}, Robert and {Nakano}, Toshio and {Castro}, Daniel and {Enoto}, Teruaki and {Hiraga}, Junko S. and {Hughes}, John P. and {Maeda}, Yoshitomo and {Nobukawa}, Masayoshi and {Safi-Harb}, Samar and {Slane}, Patrick O. and {Smith}, Randall K. and {Uchida}, Hiroyuki},
        title = "{Discriminating the Progenitor Type of Supernova Remnants with Iron K-shell Emission}",
      journal = {\apjl},
         year = 2014,
        month = apr,
       volume = {785},
       number = {2},
          eid = {L27},
        pages = {L27},
          doi = {10.1088/2041-8205/785/2/L27},
archivePrefix = {arXiv},
       eprint = {1403.5154},
 primaryClass = {astro-ph.HE},
       adsurl = {https://ui.adsabs.harvard.edu/abs/2014ApJ...785L..27Y}
}

@ARTICLE{Celotti2001,
       author = {{Celotti}, Annalisa and {Ghisellini}, Gabriele and {Chiaberge}, Marco},
        title = "{Large-scale jets in active galactic nuclei: multiwavelength mapping}",
      journal = {\mnras},
         year = 2001,
        month = feb,
       volume = {321},
       number = {1},
        pages = {L1-L5},
          doi = {10.1046/j.1365-8711.2001.04160.x},
archivePrefix = {arXiv},
       eprint = {astro-ph/0008021},
 primaryClass = {astro-ph},
       adsurl = {https://ui.adsabs.harvard.edu/abs/2001MNRAS.321L...1C}
}

@ARTICLE{Abbott2017,
       author = {{Abbott}, B.~P. and {Abbott}, R. and {Abbott}, T.~D. and {Acernese}, F. and {Ackley}, K. and {Adams}, C. and {Adams}, T. and {Addesso}, P. and {Adhikari}, R.~X. and {Adya}, V.~B. and {Affeldt}, C. and {Afrough}, M. and {Agarwal}, B. and {Agathos}, M. and {Agatsuma}, K. and {Aggarwal}, N. and {Aguiar}, O.~D. and {Aiello}, L. and {Ain}, A. and {Ajith}, P. and {Allen}, B. and {Allen}, G. and {Allocca}, A. and {Altin}, P.~A. and {Amato}, A. and {Ananyeva}, A. and {Anderson}, S.~B. and {Anderson}, W.~G. and {Angelova}, S.~V. and {Antier}, S. and {Appert}, S. and {Arai}, K. and {Araya}, M.~C. and {Areeda}, J.~S. and {Arnaud}, N. and {Arun}, K.~G. and {Ascenzi}, S. and {Ashton}, G. and {Ast}, M. and {Aston}, S.~M. and {Astone}, P. and {Atallah}, D.~V. and {Aufmuth}, P. and {Aulbert}, C. and {AultONeal}, K. and {Austin}, C. and {Avila-Alvarez}, A. and {Babak}, S. and {Bacon}, P. and {Bader}, M.~K.~M. and {Bae}, S. and {Bailes}, M. and {Baker}, P.~T. and {Baldaccini}, F. and {Ballardin}, G. and {Ballmer}, S.~W. and {Banagiri}, S. and {Barayoga}, J.~C. and {Barclay}, S.~E. and {Barish}, B.~C. and {Barker}, D. and {Barkett}, K. and {Barone}, F. and {Barr}, B. and {Barsotti}, L. and {Barsuglia}, M. and {Barta}, D. and {Barthelmy}, S.~D. and {Bartlett}, J. and {Bartos}, I. and {Bassiri}, R. and {Basti}, A. and {Batch}, J.~C. and {Bawaj}, M. and {Bayley}, J.~C. and {Bazzan}, M. and {B{\'e}csy}, B. and {Beer}, C. and {Bejger}, M. and {Belahcene}, I. and {Bell}, A.~S. and {Berger}, B.~K. and {Bergmann}, G. and {Bernuzzi}, S. and {Bero}, J.~J. and {Berry}, C.~P.~L. and {Bersanetti}, D. and {Bertolini}, A. and {Betzwieser}, J. and {Bhagwat}, S. and {Bhandare}, R. and {Bilenko}, I.~A. and {Billingsley}, G. and {Billman}, C.~R. and {Birch}, J. and {Birney}, R. and {Birnholtz}, O. and {Biscans}, S. and {Biscoveanu}, S. and {Bisht}, A. and {Bitossi}, M. and {Biwer}, C. and {Bizouard}, M.~A. and {Blackburn}, J.~K. and {Blackman}, J. and {Blair}, C.~D. and {Blair}, D.~G. and {Blair}, R.~M. and {Bloemen}, S. and {Bock}, O. and {Bode}, N. and {Boer}, M. and {Bogaert}, G. and {Bohe}, A. and {Bondu}, F. and {Bonilla}, E. and {Bonnand}, R. and {Boom}, B.~A. and {Bork}, R. and {Boschi}, V. and {Bose}, S. and {Bossie}, K. and {Bouffanais}, Y. and {Bozzi}, A. and {Bradaschia}, C. and {Brady}, P.~R. and {Branchesi}, M. and {Brau}, J.~E. and {Briant}, T. and {Brillet}, A. and {Brinkmann}, M. and {Brisson}, V. and {Brockill}, P. and {Broida}, J.~E. and {Brooks}, A.~F. and {Brown}, D.~A. and {Brown}, D.~D. and {Brunett}, S. and {Buchanan}, C.~C. and {Buikema}, A. and {Bulik}, T. and {Bulten}, H.~J. and {Buonanno}, A. and {Buskulic}, D. and {Buy}, C. and {Byer}, R.~L. and {Cabero}, M. and {Cadonati}, L. and {Cagnoli}, G. and {Cahillane}, C. and {Calder{\'o}n Bustillo}, J. and {Callister}, T.~A. and {Calloni}, E. and {Camp}, J.~B. and {Canepa}, M. and {Canizares}, P. and {Cannon}, K.~C. and {Cao}, H. and {Cao}, J. and {Capano}, C.~D. and {Capocasa}, E. and {Carbognani}, F. and {Caride}, S. and {Carney}, M.~F. and {Carullo}, G. and {Casanueva Diaz}, J. and {Casentini}, C. and {Caudill}, S. and {Cavagli{\`a}}, M. and {Cavalier}, F. and {Cavalieri}, R. and {Cella}, G. and {Cepeda}, C.~B. and {Cerd{\'a}-Dur{\'a}n}, P. and {Cerretani}, G. and {Cesarini}, E. and {Chamberlin}, S.~J. and {Chan}, M. and {Chao}, S. and {Charlton}, P. and {Chase}, E. and {Chassande-Mottin}, E. and {Chatterjee}, D. and {Chatziioannou}, K. and {Cheeseboro}, B.~D. and {Chen}, H.~Y. and {Chen}, X. and {Chen}, Y. and {Cheng}, H. -P. and {Chia}, H. and {Chincarini}, A. and {Chiummo}, A. and {Chmiel}, T. and {Cho}, H.~S. and {Cho}, M. and {Chow}, J.~H. and {Christensen}, N. and {Chu}, Q. and {Chua}, A.~J.~K. and {Chua}, S.},
        title = "{GW170817: Observation of Gravitational Waves from a Binary Neutron Star Inspiral}",
      journal = {\prl},
         year = 2017,
        month = oct,
       volume = {119},
       number = {16},
          eid = {161101},
        pages = {161101},
          doi = {10.1103/PhysRevLett.119.161101},
archivePrefix = {arXiv},
       eprint = {1710.05832},
 primaryClass = {gr-qc},
       adsurl = {https://ui.adsabs.harvard.edu/abs/2017PhRvL.119p1101A}
}

@ARTICLE{Abbott2017-2,
       author = {{Abbott}, B.~P. and {Abbott}, R. and {Abbott}, T.~D. and {Acernese}, F. and {Ackley}, K. and {Adams}, C. and {Adams}, T. and {Addesso}, P. and {Adhikari}, R.~X. and {Adya}, V.~B. and {Affeldt}, C. and {Afrough}, M. and {Agarwal}, B. and {Agathos}, M. and {Agatsuma}, K. and {Aggarwal}, N. and {Aguiar}, O.~D. and {Aiello}, L. and {Ain}, A. and {Ajith}, P. and {Allen}, B. and {Allen}, G. and {Allocca}, A. and {Altin}, P.~A. and {Amato}, A. and {Ananyeva}, A. and {Anderson}, S.~B. and {Anderson}, W.~G. and {Angelova}, S.~V. and {Antier}, S. and {Appert}, S. and {Arai}, K. and {Araya}, M.~C. and {Areeda}, J.~S. and {Arnaud}, N. and {Arun}, K.~G. and {Ascenzi}, S. and {Ashton}, G. and {Ast}, M. and {Aston}, S.~M. and {Astone}, P. and {Atallah}, D.~V. and {Aufmuth}, P. and {Aulbert}, C. and {AultONeal}, K. and {Austin}, C. and {Avila-Alvarez}, A. and {Babak}, S. and {Bacon}, P. and {Bader}, M.~K.~M. and {Bae}, S. and {Baker}, P.~T. and {Baldaccini}, F. and {Ballardin}, G. and {Ballmer}, S.~W. and {Banagiri}, S. and {Barayoga}, J.~C. and {Barclay}, S.~E. and {Barish}, B.~C. and {Barker}, D. and {Barkett}, K. and {Barone}, F. and {Barr}, B. and {Barsotti}, L. and {Barsuglia}, M. and {Barta}, D. and {Barthelmy}, S.~D. and {Bartlett}, J. and {Bartos}, I. and {Bassiri}, R. and {Basti}, A. and {Batch}, J.~C. and {Bawaj}, M. and {Bayley}, J.~C. and {Bazzan}, M. and {B{\'e}csy}, B. and {Beer}, C. and {Bejger}, M. and {Belahcene}, I. and {Bell}, A.~S. and {Berger}, B.~K. and {Bergmann}, G. and {Bero}, J.~J. and {Berry}, C.~P.~L. and {Bersanetti}, D. and {Bertolini}, A. and {Betzwieser}, J. and {Bhagwat}, S. and {Bhandare}, R. and {Bilenko}, I.~A. and {Billingsley}, G. and {Billman}, C.~R. and {Birch}, J. and {Birney}, R. and {Birnholtz}, O. and {Biscans}, S. and {Biscoveanu}, S. and {Bisht}, A. and {Bitossi}, M. and {Biwer}, C. and {Bizouard}, M.~A. and {Blackburn}, J.~K. and {Blackman}, J. and {Blair}, C.~D. and {Blair}, D.~G. and {Blair}, R.~M. and {Bloemen}, S. and {Bock}, O. and {Bode}, N. and {Boer}, M. and {Bogaert}, G. and {Bohe}, A. and {Bondu}, F. and {Bonilla}, E. and {Bonnand}, R. and {Boom}, B.~A. and {Bork}, R. and {Boschi}, V. and {Bose}, S. and {Bossie}, K. and {Bouffanais}, Y. and {Bozzi}, A. and {Bradaschia}, C. and {Brady}, P.~R. and {Branchesi}, M. and {Brau}, J.~E. and {Briant}, T. and {Brillet}, A. and {Brinkmann}, M. and {Brisson}, V. and {Brockill}, P. and {Broida}, J.~E. and {Brooks}, A.~F. and {Brown}, D.~A. and {Brown}, D.~D. and {Brunett}, S. and {Buchanan}, C.~C. and {Buikema}, A. and {Bulik}, T. and {Bulten}, H.~J. and {Buonanno}, A. and {Buskulic}, D. and {Buy}, C. and {Byer}, R.~L. and {Cabero}, M. and {Cadonati}, L. and {Cagnoli}, G. and {Cahillane}, C. and {Calder{\'o}n Bustillo}, J. and {Callister}, T.~A. and {Calloni}, E. and {Camp}, J.~B. and {Canepa}, M. and {Canizares}, P. and {Cannon}, K.~C. and {Cao}, H. and {Cao}, J. and {Capano}, C.~D. and {Capocasa}, E. and {Carbognani}, F. and {Caride}, S. and {Carney}, M.~F. and {Casanueva Diaz}, J. and {Casentini}, C. and {Caudill}, S. and {Cavagli{\`a}}, M. and {Cavalier}, F. and {Cavalieri}, R. and {Cella}, G. and {Cepeda}, C.~B. and {Cerd{\'a}-Dur{\'a}n}, P. and {Cerretani}, G. and {Cesarini}, E. and {Chamberlin}, S.~J. and {Chan}, M. and {Chao}, S. and {Charlton}, P. and {Chase}, E. and {Chassande-Mottin}, E. and {Chatterjee}, D. and {Chatziioannou}, K. and {Cheeseboro}, B.~D. and {Chen}, H.~Y. and {Chen}, X. and {Chen}, Y. and {Cheng}, H. -P. and {Chia}, H. and {Chincarini}, A. and {Chiummo}, A. and {Chmiel}, T. and {Cho}, H.~S. and {Cho}, M. and {Chow}, J.~H. and {Christensen}, N. and {Chu}, Q. and {Chua}, A.~J.~K. and {Chua}, S. and {Chung}, A.~K.~W. and {Chung}, S. and {Ciani}, G.},
        title = "{Multi-messenger Observations of a Binary Neutron Star Merger}",
      journal = {\apjl},
         year = 2017,
        month = oct,
       volume = {848},
       number = {2},
          eid = {L12},
        pages = {L12},
          doi = {10.3847/2041-8213/aa91c9},
archivePrefix = {arXiv},
       eprint = {1710.05833},
 primaryClass = {astro-ph.HE},
       adsurl = {https://ui.adsabs.harvard.edu/abs/2017ApJ...848L..12A}
}

@ARTICLE{Troja2020,
       author = {{Troja}, E. and {van Eerten}, H. and {Zhang}, B. and {Ryan}, G. and {Piro}, L. and {Ricci}, R. and {O'Connor}, B. and {Wieringa}, M.~H. and {Cenko}, S.~B. and {Sakamoto}, T.},
        title = "{A thousand days after the merger: Continued X-ray emission from GW170817}",
      journal = {\mnras},
         year = 2020,
        month = nov,
       volume = {498},
       number = {4},
        pages = {5643-5651},
          doi = {10.1093/mnras/staa2626},
archivePrefix = {arXiv},
       eprint = {2006.01150},
 primaryClass = {astro-ph.HE},
       adsurl = {https://ui.adsabs.harvard.edu/abs/2020MNRAS.498.5643T}
}

@ARTICLE{Ren2022,
       author = {{Ren}, J. and {Dai}, Z.~G.},
        title = "{Broad-band emission from a kilonova ejecta-pulsar wind Nebula system: late-time X-ray afterglow rebrightening of GRB 170817A}",
      journal = {\mnras},
         year = 2022,
        month = jun,
       volume = {512},
       number = {4},
        pages = {5572-5579},
          doi = {10.1093/mnras/stac797},
archivePrefix = {arXiv},
       eprint = {2203.08576},
 primaryClass = {astro-ph.HE},
       adsurl = {https://ui.adsabs.harvard.edu/abs/2022MNRAS.512.5572R}
}

@ARTICLE{Hajela2022,
       author = {{Hajela}, A. and {Margutti}, R. and {Bright}, J.~S. and {Alexander}, K.~D. and {Metzger}, B.~D. and {Nedora}, V. and {Kathirgamaraju}, A. and {Margalit}, B. and {Radice}, D. and {Guidorzi}, C. and {Berger}, E. and {MacFadyen}, A. and {Giannios}, D. and {Chornock}, R. and {Heywood}, I. and {Sironi}, L. and {Gottlieb}, O. and {Coppejans}, D. and {Laskar}, T. and {Cendes}, Y. and {Duran}, R. Barniol and {Eftekhari}, T. and {Fong}, W. and {McDowell}, A. and {Nicholl}, M. and {Xie}, X. and {Zrake}, J. and {Bernuzzi}, S. and {Broekgaarden}, F.~S. and {Kilpatrick}, C.~D. and {Terreran}, G. and {Villar}, V.~A. and {Blanchard}, P.~K. and {Gomez}, S. and {Hosseinzadeh}, G. and {Matthews}, D.~J. and {Rastinejad}, J.~C.},
        title = "{Evidence for X-Ray Emission in Excess to the Jet-afterglow Decay 3.5 yr after the Binary Neutron Star Merger GW 170817: A New Emission Component}",
      journal = {\apjl},
         year = 2022,
        month = mar,
       volume = {927},
       number = {1},
          eid = {L17},
        pages = {L17},
          doi = {10.3847/2041-8213/ac504a},
archivePrefix = {arXiv},
       eprint = {2104.02070},
 primaryClass = {astro-ph.HE},
       adsurl = {https://ui.adsabs.harvard.edu/abs/2022ApJ...927L..17H}
}

@ARTICLE{Cash1979,
       author = {{Cash}, W.},
        title = "{Parameter estimation in astronomy through application of the likelihood ratio.}",
      journal = {\apj},
         year = 1979,
        month = mar,
       volume = {228},
        pages = {939-947},
          doi = {10.1086/156922},
       adsurl = {https://ui.adsabs.harvard.edu/abs/1979ApJ...228..939C}
}

@misc{MCPlot,
    author={{Lansey} Jonathan C.},
    publisher={MATLAB Central File Exchange},
    title={Plot and compare histograms; pretty by default},
    year={2025},
    url={https://www.mathworks.com/matlabcentral/fileexchange/27388-plot-and-compare-histograms-pretty-by-default}
}

@misc{200cm,
    author={{Zhaoxu Liu / slandarer}},
    publisher={MATLAB Central File Exchange},
    title={200 colormap},
    year={2025},
    url={https://www.mathworks.com/matlabcentral/fileexchange/120088-200-colormap}
}

@ARTICLE{2020Martinez,
       author = {{Mart{\'\i}nez-Rodr{\'\i}guez}, H{\'e}ctor and {Lopez}, Laura A. and {Auchettl}, Katie and {Badenes}, Carles and {Holland-Ashford}, Tyler and {Patnaude}, Daniel J. and {Lee}, Shiu-Hang and {Foster}, Adam R. and {Slane}, Patrick O.},
        title = "{Evidence of a Type Ia Progenitor for Supernova Remnant 3C 397}",
      journal = {arXiv e-prints},
         year = 2020,
        month = jun,
          eid = {arXiv:2006.08681},
        pages = {arXiv:2006.08681},
          doi = {10.48550/arXiv.2006.08681},
archivePrefix = {arXiv},
       eprint = {2006.08681},
 primaryClass = {astro-ph.HE},
       adsurl = {https://ui.adsabs.harvard.edu/abs/2020arXiv200608681M}
}

@ARTICLE{2020H,
       author = {{Hell}, Natalie and {Beiersdorfer}, Peter and {Brown}, Gregory V. and {Eckart}, Megan E. and {Kelley}, Richard L. and {Kilbourne}, Caroline A. and {Leutenegger}, Maurice A. and {Lockard}, Thomas E. and {Porter}, F. Scott and {Wilms}, J{\"o}rn},
        title = "{Highly charged ions in a new era of high resolution X‑ray astrophysics}",
      journal = {X-ray Spectrometry},
         year = 2020,
        month = jan,
       volume = {49},
       number = {1},
        pages = {218-233},
          doi = {10.1002/xrs.3107},
       adsurl = {https://ui.adsabs.harvard.edu/abs/2020XRS....49..218H}
}

@ARTICLE{2014A&A...564A.125B,
       author = {{Buchner}, J. and {Georgakakis}, A. and {Nandra}, K. and {Hsu}, L. and {Rangel}, C. and {Brightman}, M. and {Merloni}, A. and {Salvato}, M. and {Donley}, J. and {Kocevski}, D.},
        title = "{X-ray spectral modelling of the AGN obscuring region in the CDFS: Bayesian model selection and catalogue}",
      journal = {\aap},
         year = 2014,
        month = apr,
       volume = {564},
          eid = {A125},
        pages = {A125},
          doi = {10.1051/0004-6361/201322971},
archivePrefix = {arXiv},
       eprint = {1402.0004},
 primaryClass = {astro-ph.HE},
       adsurl = {https://ui.adsabs.harvard.edu/abs/2014A&A...564A.125B}
}

@ARTICLE{2021JOSS....6.3001B,
       author = {{Buchner}, Johannes},
        title = "{UltraNest - a robust, general purpose Bayesian inference engine}",
      journal = {The Journal of Open Source Software},
         year = 2021,
        month = apr,
       volume = {6},
       number = {60},
          eid = {3001},
        pages = {3001},
          doi = {10.21105/joss.03001},
archivePrefix = {arXiv},
       eprint = {2101.09604},
 primaryClass = {stat.CO},
       adsurl = {https://ui.adsabs.harvard.edu/abs/2021JOSS....6.3001B}
}

@ARTICLE{2009MNRAS.398.1601F,
       author = {{Feroz}, F. and {Hobson}, M.~P. and {Bridges}, M.},
        title = "{MULTINEST: an efficient and robust Bayesian inference tool for cosmology and particle physics}",
      journal = {\mnras},
         year = 2009,
        month = oct,
       volume = {398},
       number = {4},
        pages = {1601-1614},
          doi = {10.1111/j.1365-2966.2009.14548.x},
archivePrefix = {arXiv},
       eprint = {0809.3437},
 primaryClass = {astro-ph},
       adsurl = {https://ui.adsabs.harvard.edu/abs/2009MNRAS.398.1601F}
}

@ARTICLE{2023StSur..17..169B,
       author = {{Buchner}, Johannes},
        title = "{Nested Sampling Methods}",
      journal = {Statistics Surveys},
         year = 2023,
        month = jan,
       volume = {17},
        pages = {169-215},
          doi = {10.1214/23-SS144},
archivePrefix = {arXiv},
       eprint = {2101.09675},
 primaryClass = {stat.CO},
       adsurl = {https://ui.adsabs.harvard.edu/abs/2023StSur..17..169B}
}

@ARTICLE{2017EPJC...77..761M,
       author = {{Martinez}, Gregory D. and {McKay}, James and {Farmer}, Ben and {Scott}, Pat and {Roebber}, Elinore and {Putze}, Antje and {Conrad}, Jan},
        title = "{Comparison of statistical sampling methods with ScannerBit, the GAMBIT scanning module}",
      journal = {European Physical Journal C},
         year = 2017,
        month = nov,
       volume = {77},
       number = {11},
          eid = {761},
        pages = {761},
          doi = {10.1140/epjc/s10052-017-5274-y},
archivePrefix = {arXiv},
       eprint = {1705.07959},
 primaryClass = {hep-ph},
       adsurl = {https://ui.adsabs.harvard.edu/abs/2017EPJC...77..761M}
}

@misc{passmark5995wx,
  author       = {{PassMark Software}},
  year         = {2025},
  title        = {CPU Benchmarks: {AMD Ryzen Threadripper PRO 5995WX}},
  howpublished = {\url{https://www.cpubenchmark.net/cpu.php?cpu=AMD+Ryzen+Threadripper+PRO+5995WX}},
  note         = {Accessed 2025-12-16}
}
\bibliographystyle{aasjournalv7}

\end{document}